\DocumentMetadata{
  lang        = en-US,
  pdfversion  = 2.0,
  pdfstandard = ua-1
}

\documentclass[sigconf]{acmart}
\usepackage{xcolor}
\usepackage{booktabs} 
\usepackage{amsmath}
\usepackage{graphicx}
\usepackage{multirow}
\usepackage{subfig}
\usepackage{float}
\usepackage{placeins}
\usepackage{tabularx}
\usepackage{array}
\usepackage{makecell}
\newcolumntype{Y}{>{\raggedright\arraybackslash}X}

\AtBeginDocument{%
  }

\copyrightyear{2026}
\acmYear{2026}
\setcopyright{cc}
\setcctype{by}
\acmConference[ASSETS '26]{The 28th International ACM SIGACCESS Conference on Computers and Accessibility}{October 25--28, 2026}{Vila Nova de Gaia, Portugal}
\acmBooktitle{The 28th International ACM SIGACCESS Conference on Computers and Accessibility (ASSETS '26), October 25--28, 2026, Vila Nova de Gaia, Portugal}
\acmDOI{10.1145/3797867.3829017}
\acmISBN{979-8-4007-2521-0/2026/10}

\hypersetup{
  pdftitle            = {Balancing Safety and Autonomy: Accessibility-Oriented Interventions in Generative AI for Cognitive Impairment},
  pdfauthor           = {Yibo Meng, Jingruo Chen, Lyumanshan Ye, Bingyi Liu, Zhicong Lu},
  pdfsubject          = {ACM SIGACCESS submission},
  pdfkeywords         = {generative AI, cognitive impairment, older adults, accessibility, autonomy, safety, human oversight, dementia, AI-assisted care, human-computer interaction},
  pdflang             = en-US,
  pdfdisplaydoctitle  = true
}

\begin{document}
\title{Balancing Safety and Autonomy: Accessibility-Oriented Interventions in Generative AI for Cognitive Impairment}

\author{Yibo Meng}
\authornote{Both authors contributed equally to this research.}
\affiliation{%
  \institution{Weill Cornell Medicine}
  \city{New York}
  \country{United States}
}
  \email{yim4007@med.cornell.edu}

\author{Jingruo Chen}
\authornotemark[1]
\affiliation{
  \institution{Information Science, Cornell University}
  \city{Ithaca}
  \state{New York}
  \country{United States}
}
\email{jc3564@cornell.edu}

\author{Lyumanshan Ye}
\affiliation{
  \institution{Shanghai Jiao Tong University}
  \city{Shanghai}
  \country{China}
}

\author{Bingyi Liu}
\affiliation{
  \institution{University of Michigan}
  \city{Ann Arbor}
  \state{Michigan}
  \country{United States}
}
\email{bingyi@umich.edu}

\author{Zhicong Lu}
\authornote{Corresponding author.}
\affiliation{
  \institution{Department of Computer Science, George Mason University}
  \city{Fairfax}
  \state{Virginia}
  \country{United States}
}
\email{zlu6@gmu.edu}

\begin{abstract}
Generative AI systems are increasingly used by older adults with cognitive impairment for everyday tasks such as information seeking, health management, and communication. While these systems provide flexible, language-based support, their open-ended outputs introduce risks of over-reliance, misinterpretation, and inappropriate decision-making. Prior work has focused on usability and adoption, with limited attention to how system design shapes users’ participation in decision-making and the distribution of agency in care contexts. We present a qualitative study of 45 individuals with cognitive impairment and their caregivers. We identify five accessibility-oriented mechanisms: AI Capability Constraint, Human Oversight Embedding, Cognitive Engagement Maintenance, Human–AI Relationship Regulation, and Risk Transparency and Control, through which systems structure interaction. These mechanisms both support and constrain users by redistributing decision-making across users and caregivers. We show that their effects vary by impairment level: while protective mechanisms support users with severe impairment, they can restrict autonomy for those with mild impairment. As impairment progresses, tensions become less visible as user participation diminishes. Our findings highlight the need for dynamic designs that balance safety and autonomy in AI-supported care.
\end{abstract}



\maketitle

\hypersetup{
  pdfsubject  = {ACM SIGACCESS submission},
  pdfkeywords = {generative AI, cognitive impairment, older adults, accessibility, autonomy, safety, human oversight, dementia, AI-assisted care, human-computer interaction}
}

\section{Introduction}

Generative AI systems are increasingly integrated into everyday technologies such as mobile applications, conversational assistants, and digital health tools, where they are used for tasks including information seeking, communication, and decision support~\cite{treder2024introduction, arets2026shared, breithaupt2025designing}. For older adults with cognitive impairment, particularly those experiencing conditions such as dementia, these systems are encountered in contexts such as medication management, symptom inquiry, and daily coordination~\cite{hou2025using, qi2022artificial, wang2024enhancing}. Compared to earlier rule-based or task-specific assistive technologies, generative AI enables flexible, language-based interaction and can be applied across a wide range of everyday situations~\cite{kot2026exploring}. At the same time, its open-ended and probabilistic outputs introduce new challenges: responses may be difficult to verify, and users may rely on them without fully understanding their limitations.

Prior HCI research has addressed cognitive impairment through technologies that simplify interaction, externalize memory, and scaffold task completion~\cite{pradhan2023towards, hu2024designing, guedes2024scaffolding}. Conversational interfaces have further reduced barriers by enabling natural language interaction~\cite{pradhan2020use, baldauf2018exploring}. More broadly, work on aging and AI highlights that technology use is embedded within care networks involving family members, caregivers, and healthcare professionals, where decision-making is often distributed rather than individual~\cite{vidas2024wouldn, houben2022designing, dai2021surfacing}. At the same time, researchers have emphasized tensions between support and autonomy, noting that assistive systems may both enable independence and introduce new forms of dependency or control~\cite{coghlan2021dignity, berridge2022control}.

Generative AI intensifies these tensions. Its fluent, contextually appropriate responses expand support but also increase risks of over-reliance, misinterpretation, and inappropriate decision-making~\cite{gilman2024training}. In response, systems incorporate interventions such as output constraints, human oversight, and structured interaction. These interventions do not merely reduce risk; they reshape how users participate in understanding, judgment, and action, redistributing decision-making across users, caregivers, and technologies. However, how these mechanisms operate in everyday use remains underexplored.

To address this gap, we investigate how generative AI systems are used by older adults with cognitive impairment and how accessibility-oriented interventions shape interaction in real-world settings. Drawing on semi-structured interviews with 45 individuals with cognitive impairment and their caregivers, we identify five recurring mechanisms through which systems structure use: \textit{AI Capability Constraint}, \textit{Human Oversight Embedding}, \textit{Cognitive Engagement Maintenance}, \textit{Human--AI Relationship Regulation}, and \textit{Risk Transparency and Control}. We show that these mechanisms operate through two distinct logics: \textit{enablement}, which sustains users’ participation in understanding and reasoning, and \textit{protection}, which reduces risk by constraining system behavior and redistributing decision-making.

Our findings reveal that these mechanisms produce uneven and sometimes conflicting effects across levels of cognitive impairment. While protective mechanisms provide necessary safeguards for users with severe impairment, they can constrain users with mild impairment, limiting participation and reducing perceived autonomy. At the same time, tensions between safety and autonomy do not disappear as impairment progresses; rather, they become less visible as users’ ability to perceive, articulate, or resist constraints diminishes. In many cases, apparent improvements in usability or safety are achieved through reduced user participation, shifts in expectations, or the transfer of cognitive burden to caregivers.

This work contributes to HCI and accessibility research in three ways. First, we introduce accessibility-oriented intervention mechanisms as an analytic lens for understanding how generative AI systems structure interaction beyond interface-level adaptations. Second, we demonstrate that cognitive impairment level is a key moderating factor shaping how these mechanisms are experienced, highlighting the need for dynamic and context-sensitive design. Third, we surface the relational and power dynamics embedded in AI-mediated support, showing how design decisions redistribute authority across users, caregivers, and systems, and raising important ethical considerations for the design of generative AI in aging contexts.

\section{Related Work}
\subsection{HCI Systems for Cognitive Impairment}
Prior HCI research has long explored technologies to support individuals with cognitive impairment~\cite{keates2009cognitive, dixon2022mobile, kim2024opportunities, loveleen2023explanation, khot2026temporal}, including systems for evaluation~\cite{su2024system}, wayfinding~\cite{chang2008context}, and memory assistance~\cite{pradhan2023towards}. These systems range from reminder-based applications~\cite{pradhan2023towards} and cognitive training tools~\cite{li2024effect, eisapour2018participatory} to digital and creativity support~\cite{piper2016technological, king2024safespace, lazar2016designing, vetter2026calls}. Across this body of work, a central goal has been to compensate for reduced memory, attention, and executive function by simplifying interaction and externalizing cognitive processes~\cite{boehmer2025too, dixon2024investigating, dixon2020role}.

Empirical studies show that individuals with cognitive impairment face challenges in interpreting complex information, maintaining context across interactions, and navigating multi-step tasks in everyday technologies such as smartphones~\cite{dixon2022mobile}. For instance, prior work has found that people with mild-to-moderate dementia use mobile devices for medication management and health tracking~\cite{dixon2021taking}, yet struggle with app navigation and task execution under stress~\cite{vines2015age, pradhan2020use}. Voice-based assistants have been explored as an alternative interaction modality, as conversational interfaces reduce the need for technical expertise and enable more natural interaction~\cite{pradhan2020use, baldauf2018exploring, qiu2025voice}; however, reliability concerns, particularly for reminders and memory support, limit sustained adoption.

In response, design approaches have emphasized structured guidance, repetition, reduced information density, and scaffolding to facilitate understanding and task completion~\cite{hu2024designing, guedes2024scaffolding, huang2026collaborative, hu2026looking}. At the same time, research highlights broader socio-technical concerns, including autonomy, privacy, and trust~\cite{madjaroff2017narratives, mentis2019upside}. For instance, studies of assistive technologies and social robots point to tensions between support and dignity, where users worry about losing independence or becoming overly reliant on automated assistance~\cite{coghlan2021dignity}. Emerging directions explore more personalized and context-aware systems, such as co-designed solutions that adapt to users’ abilities~\cite{wang2025enabling, kim2022mymove, ppali2025creating, bircanin2021including} and immersive environments that incorporate social and sensory context~\cite{flynn2025multi, xygkou2024mindtalker}.

With the emergence of generative AI systems, these challenges are further complicated by the open-ended and probabilistic nature of system outputs, but this same flexibility also creates opportunities to dynamically tailor prompts, explanations, and decision support to users’ situational context and cognitive capacity~\cite{kot2026exploring, treder2024introduction, arets2026shared, breithaupt2025designing}. Unlike earlier rule-based or task-specific tools, generative AI systems can produce fluent, contextually appropriate responses that may be difficult for users to critically assess~\cite{gilman2024training}. While existing research has begun to examine the use of conversational AI in cognitive support contexts~\cite{saha2025ai, smriti2024emotion}, less attention has been given to how these systems shape users’ understanding, decision-making processes, and reliance patterns in practice. Our work builds on this literature by examining how generative AI systems are embedded in real-world use and how their behaviors structure user interaction.

\subsection{Generative AI and the Aging Population}
HCI research has examined the adoption and use of AI systems among older adults, highlighting both opportunities and challenges associated with aging populations~\cite{mathur2025feels, huang2025designing}. Generative AI systems, particularly conversational agents, are increasingly positioned as tools for information access, companionship, and everyday support~\cite{huang2025designing, mccarren2026exploring, vinay2025grace, shi2026humans, mingxi2025can}. Prior studies suggest that such systems can lower barriers to technology use by enabling natural language interaction and providing on-demand assistance, which is especially beneficial for users with limited technical experience~\cite{ishibashi2025preference}.

At the same time, research on aging and technology has emphasized the importance of maintaining autonomy, dignity, and independence in later life~\cite{berridge2022control, coghlan2021dignity}. Tanis and Lewis further invoke the \textit{dignity of risk} to argue that people with cognitive disabilities should retain opportunities to make choices and accept reasonable risks rather than being categorically shielded from them~\cite{10.1145/3386296.3386303}. This perspective highlights how protective interventions may reduce harm while also restricting users' autonomy and participation. While AI systems can support decision-making and reduce effort, they may also introduce new forms of dependency, particularly when users rely on system outputs without fully understanding their basis~\cite{shandilya2022understanding, surani2026co}. This concern is amplified in contexts where cognitive abilities are declining, as users may have difficulty evaluating information quality or recognizing uncertainty~\cite{kim2026clarifying, tang2025ai}.

Prior work further shows that technology use in aging contexts is embedded within broader care networks involving family members, caregivers, and healthcare professionals~\cite{vidas2024wouldn, houben2022designing, dai2021surfacing, smriti2024emotion}. These actors often participate in or mediate technology use, contributing to distributed or shared decision-making processes~\cite{weng2026blessing, gui2023enhancing}. For example, studies of dementia care demonstrate that individuals frequently rely on socio-technical arrangements combining digital tools and human support to maintain independence while managing risks~\cite{li2023any}. Similarly, research on AI-enabled cognitive support systems highlights the importance of collaborative oversight, where caregivers play an active role in monitoring and interpreting system outputs~\cite{akter2024examining}.

Recent work on generative AI specifically points to emerging tensions between personalization, trust, and safety~\cite{mathur2026sometimes, mathur2026wants}. Studies of AI literacy among older adults show that while users are curious about generative systems and adopt them for tasks such as translation or creative exploration, they remain concerned about misinformation, scams, and loss of control~\cite{ko2025we}. In summary, existing research has identified both the benefits of AI support and the risks of over-reliance and reduced autonomy. However, these dimensions are often treated as parallel considerations rather than as fundamentally interconnected design tensions. In this work, we examine how generative AI systems simultaneously support and constrain users, revealing how safety-oriented design choices can reshape autonomy in practice, particularly across different stages of cognitive impairment in the aging population.

\section{Method}

This study adopts a qualitative research design to examine how individuals with cognitive impairment interact with accessible generative AI systems in everyday life. Semi-structured interviews were used to capture participants’ lived experiences, focusing on the systems they encounter, how these systems operate in specific contexts, and how interaction processes shape understanding and decision-making.
For participants whose clinical assessments indicated that direct interviews were not feasible due to the severity of their condition, caregivers were recruited as proxy informants. These proxy accounts were intended to capture observable usage patterns and to reflect caregivers' own understanding of participants' AI use based on sustained involvement in daily care, rather than to substitute for participants' subjective experiences.

For the purposes of this study, we define "generative AI systems" as systems that support the everyday health management and social support needs of older adults with cognitive impairment primarily through natural language generation and conversational interaction. This definition encompasses LLM-based medical question-answering assistants, cognitive training tools with multi-turn dialogue capabilities, intelligent health management applications, and conversational interfaces for emotional companionship. Although these systems vary in underlying technical architectures, they share natural language interaction as their primary interface modality. Our analysis focuses on their shared interaction logics rather than differences in technical implementation, which is warranted by our finding that the five mechanisms manifest across multiple system types rather than being confined to any single category.

\subsection{Participants}
\label{sec:participants}

A total of 45 individuals with cognitive impairment were included in the study, comprising 31 direct participants (P1–P31) and 14 represented through caregiver interviews (G1–G14, where G denotes caregiver/guardian). Based on clinical assessment, participants were categorized as mild (n=22), moderate (n=12), and severe (n=11). The sample spanned both urban and rural contexts, with education levels ranging from semi-literate to bachelor's degree (Table~\ref{tab:patients}).

\begin{table*}[!ht]
\centering
\caption{Participant Demographics (Patients). U/R: Urban/Rural.}
\label{tab:patients}
\begin{tabular}{l l l l l l l}
\toprule
Patient ID & Severity & Role & Age & Gender & U/R & Education \\
\midrule
P1 & Mild & Patient & 65 & F & R & Junior high school \\
P2 & Mild & Patient & 66 & F & U & Primary school \\
P3 & Mild & Patient & 67 & M & R & Junior high school \\
P4 & Mild & Patient & 70 & M & U & Junior high school \\
P5 & Mild & Patient & 70 & M & R & Bachelor \\
P6 & Mild & Patient & 70 & F & U & Primary school \\
P7 & Mild & Patient & 69 & M & R & Semi-literate \\
P8 & Mild & Patient & 66 & M & U & Primary school \\
P9 & Mild & Patient & 66 & M & R & Primary school \\
P10 & Mild & Patient & 67 & F & R & Junior high school \\
P11 & Mild & Patient & 60 & M & U & Junior high school \\
P12 & Mild & Patient & 73 & F & R & Junior high school \\
P13 & Mild & Patient & 72 & M & U & Junior high school \\
P14 & Mild & Patient & 67 & F & U & Junior high school \\
P15 & Mild & Patient & 65 & M & R & Primary school \\
P16 & Mild & Patient & 66 & F & U & Junior high school \\
P17 & Mild & Patient & 68 & F & R & Junior high school \\
P18 & Mild & Patient & 71 & F & R & Primary school \\
P19 & Mild & Patient & 72 & F & R & Primary school \\
P20 & Mild & Patient & 77 & M & U & Primary school \\
P21 & Mild & Patient & 72 & F & U & Semi-literate \\
P22 & Mild & Patient & 71 & F & R & Semi-literate \\
P23 & Moderate & Patient & 69 & M & U & Primary school \\
P24 & Moderate & Patient & 69 & M & R & Primary school \\
P25 & Moderate & Patient & 67 & M & U & Junior high school \\
P26 & Moderate & Patient & 68 & F & R & Junior high school \\
P27 & Moderate & Patient & 68 & F & U & Primary school \\
P28 & Moderate & Patient & 69 & M & U & Junior high school \\
P29 & Moderate & Patient & 64 & F & R & High school \\
P30 & Moderate & Patient & 65 & F & U & Junior high school \\
P31 & Moderate & Patient & 67 & M & R & Junior high school \\
\bottomrule
\end{tabular}
\end{table*}

Caregivers included 7 doctors and 7 family members, aged 39 to 57, all of whom had sustained involvement in patients’ daily care or medical decision-making and were familiar with their technology use.

Participants were recruited through purposive sampling from community health service centers and collaborating hospitals. Doctors first screened eligible patients based on diagnostic records and communication ability, after which the research team obtained informed consent from participants and/or their caregivers. Inclusion criteria required a clinical diagnosis with a clear severity level (via MMSE or physician assessment), age 60 or above, recent engagement with relevant technologies, and sufficient communication ability or caregiver-supported expression. Caregivers were required to have stable relationships with participants and knowledge of their technology use. Individuals unable to communicate, with severe psychiatric or neurological conditions affecting expression, or in acute medical states were excluded.

The systems encountered by participants spanned a range of applications and conversational interfaces that integrated generative AI capabilities to varying degrees, including LLM-based medical question-answering assistants, cognitive training tools with multi-turn dialogue capabilities, intelligent health management applications incorporating natural language interaction, and conversational interfaces designed for emotional companionship. These systems varied in their underlying technical architectures: some deeply integrated large language models, while others employed hybrid architectures combining rule-based components with generative AI. However, all of them relied primarily on natural language interaction as their main interface modality, and all were oriented toward the everyday health management and social support needs of older adults with cognitive impairment. For the purposes of this study, we refer to this class of systems collectively as "generative AI-integrated digital health systems," and our analysis focuses on their shared interaction logics and intervention mechanisms rather than on differences in underlying technical architecture.

Among the 45 participants, the types of systems they encountered varied widely and, in many instances, appeared in combination. Medical Q\&A assistants were the most frequently encountered system type; mentioned by approximately half of the participants, these tools were predominantly utilized by individuals seeking health-related information or guidance regarding their symptoms. Fifteen participants employed cognitive training tools featuring multi-turn conversational capabilities—tools that were often integrated into their daily rehabilitation regimens upon the recommendation of clinicians. Approximately two-thirds of the participants (n=28) utilized smart health management applications to some extent; notably, these applications were frequently used in conjunction with other types of systems. A smaller subset of participants (n=12) engaged with conversational interfaces designed for emotional companionship—a practice observed primarily among those experiencing mild functional impairments or living alone. Indeed, many participants interacted with more than one type of system across various contexts of their daily lives.
Despite the technical architectural differences among these systems, the five mechanisms identified in this study were manifested across multiple system types, rather than being confined in isolation to any single category. This cross-system pervasiveness supports the view that these mechanisms constitute inherent attributes of interaction structures, rather than merely being incidental byproducts of specific technical implementations; furthermore, it establishes a foundation for focusing our analysis on the shared interaction logic underlying these diverse systems.
\subsection{Ethics}

This study received institutional ethics approval, and all procedures followed principles of informed consent and participant protection. Prior to each interview, researchers carefully explained the study purpose, procedures, and potential risks, emphasizing that participation was voluntary and could be withdrawn at any time without consequence. Written consent was obtained from all participants; for individuals with limited cognitive capacity, consent was additionally obtained from their caregivers.

Given the vulnerability of the population, particular care was taken to minimize cognitive and communicative burden. Interviews used simplified language, step-by-step questioning, and allowed participants to proceed at their own pace. Researchers used repetition and confirmation to ensure mutual understanding and avoided pressuring participants when they were unable to respond fully. In such cases, relevant information was obtained through caregiver interviews.

All interviews were audio-recorded with consent, transcribed, and anonymized. Identifiable information was replaced with coded identifiers, and access to data was restricted to the research team. No unnecessary sensitive information was collected. These procedures aimed to ensure both data quality and the protection of participant well-being. Each participant received a 15 dollars as compensation.

\subsection{Data Collection}

Data were collected through semi-structured interviews lasting 35–55 minutes. Interviews with individuals with mild to moderate impairment served as the primary data source, while caregiver interviews supplemented cases where participants with severe impairment could not fully articulate their experiences.

Patient interviews began with open-ended questions about everyday encounters with generative AI and related systems. Participants were asked to recall what systems they used and in what situations. The interview then moved to detailed accounts of usage processes, including how systems presented information, whether they provided prompts, and what actions were required. Follow-up questions probed how participants understood system outputs, how decisions were made, and whether they relied on others during interaction. Interviews concluded with participants’ overall reflections, followed by researcher verification to ensure accuracy.

Where conditions permitted, interviews with caregivers were conducted following those with the direct study participants. For participants unable to take part in interviews directly, caregivers served as proxy informants, reconstructing observable interaction processes based on their ongoing involvement in daily care. These proxy accounts were limited to documenting system usage and behavioral patterns; they did not attempt to capture the participants' subjective experiences or evaluative feedback, nor were they regarded as an objective standard replacing the participants' own true feelings. Discrepancies between the accounts provided by caregivers and those of other participant groups were retained as valuable material for analysis. Additionally, caregivers offered reflections on the system's overall role and effectiveness, as well as how system usage evolved in tandem with changes in the participants' conditions (see Appendix~\ref{appendix:2}).

\subsection{Data Analysis}
We employed an inductive qualitative analysis approach to code interview data and derive themes~\cite{clarke2017thematic, braun2006using, braun2021thematic}. Three researchers jointly conducted the analysis, which proceeded iteratively alongside data collection.

In the initial stage, researchers independently performed open coding on interview transcripts without imposing predefined frameworks. Using line-by-line analysis, they marked key aspects of participants' narratives, including usage contexts, system behaviors, and user responses. Codes were kept close to the original data to preserve detail. The team then compared and consolidated codes through discussion, resolving disagreements and developing a shared coding framework. Using this framework, subsequent data were analyzed with the constant comparative method. New data were continuously compared against existing codes to identify either novel interaction patterns or extensions of prior ones. Emerging behaviors or response pathways were added as new categories, while recurring patterns were refined within existing codes. As analysis progressed, the team conducted axial coding to group initial codes into higher-level analytic units. This focused on how system behaviors were structured during use (e.g., information delivery, triggering conditions, and effects on user understanding and action). Through cross-participant comparison, stable patterns were identified and organized into the mechanism structure used in later analysis.

Although the thematic labels in the interview guide (Appendix~\ref{appendix:2}) bear surface resemblance to the five mechanisms identified in the analysis, this similarity does not indicate that analysis proceeded deductively. As Braun and Clarke~\cite{braun2006using, braun2021thematic} note, researchers cannot approach data free of theoretical commitments, and coding does not occur in an epistemological vacuum. The thematic labels in Appendix~\ref{appendix:2} served solely as organizational anchors for administering the interviews and were not used as a predefined codebook during analysis. Coding categories emerged inductively from the data: open codes were generated line-by-line without reference to the guide structure, and the five mechanisms were derived through axial coding that grouped these open codes into higher-level patterns based on recurring interaction structures across participants. For example, codes such as ``system redirects user to doctor,'' ``response scope narrowed to verified sources,'' and ``system declines high-risk requests'' were first generated independently, then grouped through axial coding into the pattern of output constraint, and subsequently conceptualized as the mechanism of AI Capability Constraint. This progression from open codes to axial categories to final mechanism labels reflects an inductive rather than deductive analytic pathway.

For participants who could not be interviewed directly, caregiver interviews (G1--G14) were incorporated as proxy accounts and subjected to the same coding process. These accounts were analyzed strictly for observable system use processes and interaction structures, and were compared with direct participant accounts to identify convergences and divergences. Discrepancies between the two were retained as analytically meaningful data rather than resolved by treating caregiver accounts as ground truth. Inferences regarding subjective experience, perceived autonomy, or evaluative responses to system constraints were derived only from direct participant accounts (P1--P31). Proxy statements that appeared to articulate participants' inner states beyond what could be directly observed were identified during analysis and removed or requalified accordingly.

Analysis continued concurrently with data collection. The team monitored whether new interviews introduced novel codes or patterns. After the 38th interview, no new categories emerged, and additional data only reinforced existing findings, indicating theoretical saturation. Overall, this process inductively derived stable structures of system use from participants' experiences without presupposing mechanisms, forming the basis for subsequent analysis.

\section{Findings}

\subsection{Embedding of Accessibility-Oriented Intervention Mechanisms}
\label{sec:4.1}

\begin{figure*}[t]
  \centering
  \includegraphics[width=\linewidth]{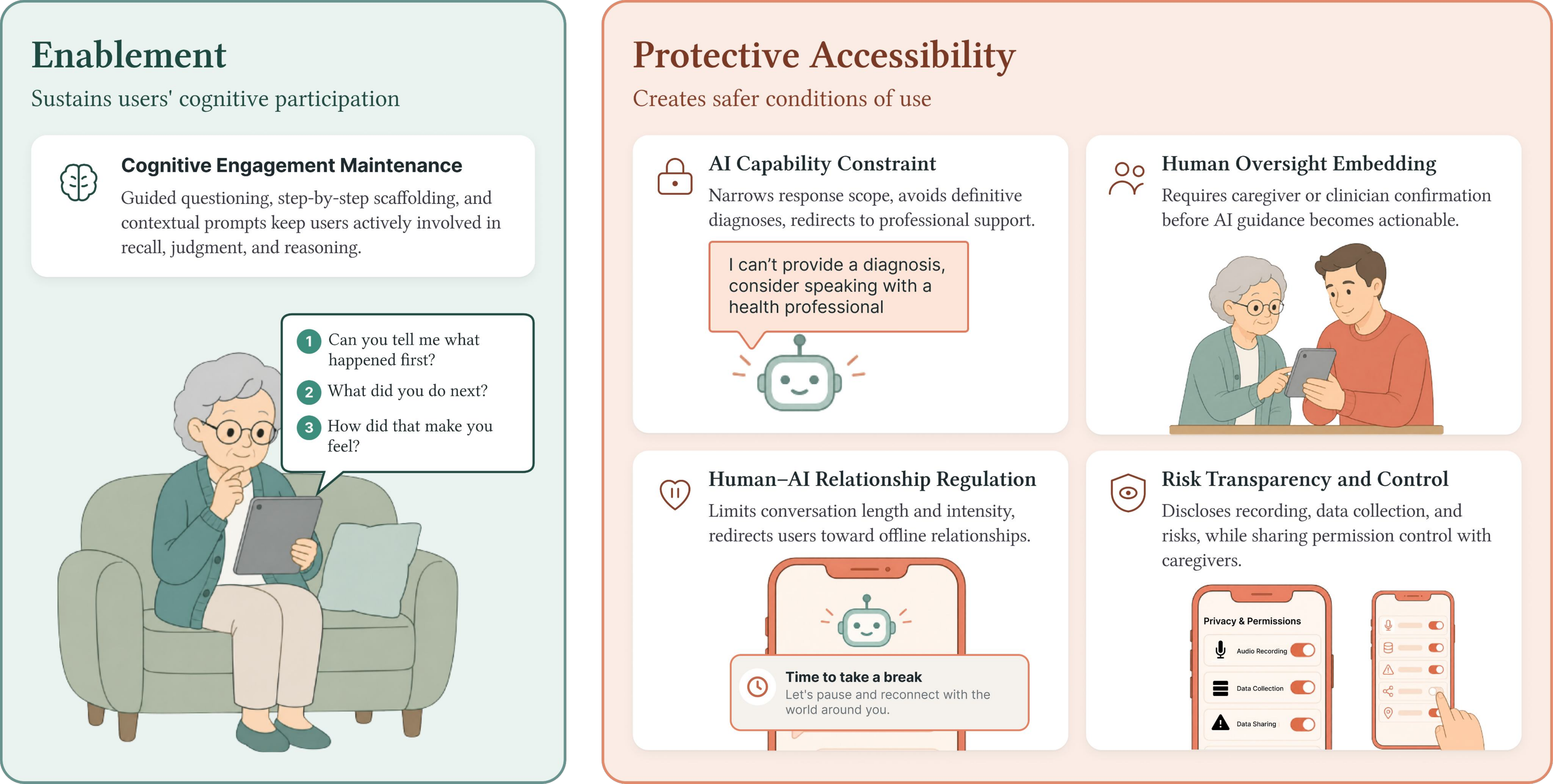}
  \Description{A two-panel diagram. The left panel, titled "Enablement, sustains users' cognitive participation," shows one mechanism: Cognitive Engagement Maintenance, illustrated by an elderly woman on a couch holding a tablet while a speech bubble asks her a sequence of three guided recall questions. The right panel, titled "Protective accessibility, creates safer conditions of use," shows four mechanisms in a two-by-two grid: AI Capability Constraint, illustrated by a chatbot declining to give a diagnosis and redirecting to a health professional; Human Oversight Embedding, illustrated by an older adult and a younger family member jointly viewing a tablet; Human--AI Relationship Regulation, illustrated by a phone notification reading "time to take a break, let's pause and reconnect with the world around you"; and Risk Transparency and Control, illustrated by two phone screens showing privacy permission toggles for audio recording, data collection, and data sharing, with a caregiver's hand adjusting the second phone.}
  \caption{Two accessibility logics underlying five intervention mechanisms in generative AI systems for older adults with cognitive impairment. \textit{Enablement} supports cognitive participation through guided scaffolding, while \textit{protective accessibility} promotes safer use by constraining AI capabilities, embedding human oversight, regulating human--AI interaction, and increasing risk transparency and user control.}
  \label{fig:mechanisms-framework}
\end{figure*}

We identify five accessibility-oriented mechanisms in generative AI systems used by older adults with cognitive impairment: \textit{AI Capability Constraint}, \textit{Human Oversight Embedding}, \textit{Cognitive Engagement Maintenance}, \textit{Human--AI Relationship Regulation}, and \textit{Risk Transparency and Control}. Figure~\ref{fig:mechanisms-framework} summarizes how these mechanisms map onto two accessibility logics, described below. We classify them as accessibility-oriented because, in the systems studied, they address barriers arising from cognitive impairment and enable users to continue participating in activities that might otherwise be difficult or unsafe. Although some mechanisms overlap with broader safety and governance approaches, their accessibility function lies in sustaining participation under conditions of reduced cognitive capacity. For example, the family-confirmation requirement encountered by P11 restricted independent action, but also allowed the activity to continue safely when unaided completion was difficult.

These mechanisms operate through two broader accessibility logics. \textit{Cognitive Engagement Maintenance} directly supports users' continued participation in understanding and judgment through step-by-step guidance, scaffolded explanations, and prompting. Rather than substituting for user cognition, it lowers barriers to comprehension and helps users remain cognitively engaged during interaction.

By contrast, \textit{AI Capability Constraint}, \textit{Human Oversight Embedding}, \textit{Human--AI Relationship Regulation}, and \textit{Risk Transparency and Control} reduce potential risks by constraining system outputs, embedding human decision-makers at critical moments, regulating the intensity of human--AI interaction, and managing the visibility and exposure of information. Together, these mechanisms reshape the conditions of use so that activities can continue under more controlled circumstances, even when independent participation is limited. We therefore conceptualize these four mechanisms as forms of \textit{protective accessibility}.

This distinction highlights a recurring tension between accessibility as \textit{enablement}, which sustains users' cognitive participation, and accessibility as \textit{protection}, which supports participation by creating safer conditions of use. The following subsections describe how each mechanism operated in participants' everyday interactions with generative AI systems.

\subsubsection{AI Capability Constraint}

Participants described systems that deliberately limited the scope or certainty of their responses, particularly in health-related contexts. Rather than offering direct diagnoses or definitive instructions, these systems often presented multiple possibilities, emphasized uncertainty, or redirected users toward professional support.

As P3 explained, \textit{``I asked whether my dizziness was serious. It didn't directly tell me what the problem was. It just said there could be many possibilities and that I should contact a doctor.''} For P3, the system's refusal to provide a definitive answer contributed to trust rather than diminishing it.

A similar pattern appeared in everyday health-management tools. P1 noted, \textit{``This app tells me what I should pay attention to when eating. It speaks more slowly and says, This is general advice.' It won't directly say, You must do this.' It always feels like an uncertain master---everything is uncertain.''} Although P1 described the system as uncertain, this uncertainty also communicated that its recommendations had limits.

Across participants' accounts, these restrictions took several forms. Systems narrowed responses to supported domains, relied on verified sources, softened expressions of certainty, or declined requests considered high risk. Medical conversational systems, for example, provided general information about diet or daily routines while avoiding direct diagnostic conclusions.

We conceptualize this recurring pattern as \textit{AI Capability Constraint}: limiting the output scope, expressive freedom, or decision-making authority of generative AI so that its behavior remains within more predictable and traceable boundaries. Such constraints may be implemented through domain-specific knowledge binding, retrieval-augmented generation, rule-based filtering, or role restrictions. The purpose is not to expand system capability, but to constrain how that capability is expressed in situations where inaccurate or overly authoritative responses may create harm.

Participants and caregivers primarily valued this pattern because it reduced exposure to authoritative-sounding but potentially inaccurate information. As G5, a physician, explained, \textit{``Sometimes the system says, `I'm not sure,' which is very good. I'm really worried that AI might say random or inappropriate things. A lot of incorrect or unsuitable information can enter users' minds this way, which is really troublesome. We need to stop it at the source.''}

These accounts show how limiting system capability could create safer and more predictable conditions for AI use. However, participants did not always experience these constraints as appropriately calibrated; we examine their implications for autonomy and practical usefulness in Section \ref{sec:4.2}.

\subsubsection{Human Oversight Embedding}
Participants also described systems that required family members, caregivers, or clinicians to become involved at key points before AI-generated information could be translated into action. Rather than allowing users to act independently on system outputs, these systems introduced additional human checkpoints into the decision process.
As P4 explained, \textit{``Once it asked me to show the content to my daughter, saying I shouldn't decide by myself. So I gave her my phone, and after she read it, she told me what to do.''} In this case, the AI-generated guidance became actionable only after it had been interpreted and approved by another person.

Participants encountered this pattern in several forms. Health-related systems sometimes advised users to contact a doctor or share information with family members. Smart health applications could send abnormal data directly to caregivers rather than displaying it only to users. Other systems required family approval before users could make changes to important settings.

P11 described one such instance: \textit{``It asked whether to change the reminder time. I thought I could just press it myself, but it said family confirmation was needed. I tried to change it, but I couldn't---it said a second person's face recognition was required.''} Here, authorization was structurally transferred from the user to a family member. Although this restricted P11's independent action, it also allowed the activity to continue under conditions in which the system treated unaided completion as potentially unsafe. AI outputs were embedded within broader family and healthcare networks rather than moving directly from system recommendation to user action. Decision pathways expanded from an individual interaction into a multi-party process involving the interpretation, confirmation, or execution of system outputs by others.

We conceptualize this recurring pattern as \textit{Human Oversight Embedding}: incorporating human participation or control into critical moments of AI use through structures that redistribute decision-making authority. Rather than relying solely on improvements to AI capability, this mechanism requires consequential decisions to be confirmed, interpreted, or taken over by caregivers, clinicians, or other responsible actors.

Caregivers and professionals valued this pattern because it provided an identifiable point of accountability and reduced the likelihood that users would act directly on inappropriate system outputs. As G7, a physician, explained, \textit{``Our users are more vulnerable than others, so hospitals often tend to introduce human intervention systems in digital technologies. At least doctors are accountable, which makes people feel much more secure. We also encourage involving family members and other caregivers.''}

These accounts show how human oversight could enable activities to continue while introducing additional safeguards and shared responsibility. At the same time, transferring decision-making authority to others could also restrict users who remained capable of making some decisions independently.

\subsubsection{Cognitive Engagement Maintenance}

Participants described systems that delayed direct answers and instead involved them in recalling information, considering options, or working through a task step by step. Rather than immediately completing tasks on users' behalf, these systems used questions and prompts to sustain participation throughout the interaction.

As P6 explained, \textit{``The app never gives me the answer immediately. It first asks whether I've eaten today, what time I ate, and lets me recall slowly. I think more slowly, but I can still remember some things.''} In this case, the system encouraged P6 to reconstruct relevant information before providing further guidance.

Similar patterns appeared across everyday activities. In dietary management, systems asked about users' current conditions before offering recommendations rather than immediately presenting a fixed plan. In memory-assistance scenarios, systems provided contextual hints instead of directly supplying an answer. P13 described one such interaction: \textit{``I forgot where I put my medicine. It asked me to think whether I had opened that drawer yesterday. I slowly remembered it step by step.''} Participants also encountered cognitive training applications that used multi-turn dialogue to support recollection, description, or simple reasoning. Across these accounts, task completion unfolded through a sequence of prompts and responses rather than through a single system output. Users remained involved in reconstructing information and forming basic judgments, while the system provided structure for progressing through the task.

We conceptualize this recurring pattern as \textit{Cognitive Engagement Maintenance}: sustaining users' participation in processes of understanding, recall, judgment, and reasoning during AI use. This mechanism commonly operated through guided questioning, incremental explanations, contextual prompts, and the decomposition of complex tasks into smaller steps.

Participants and caregivers valued this pattern because it preserved opportunities for users to exercise cognitive abilities rather than receiving every answer directly. As P21 explained, \textit{``The doctor told me to use this app, saying people need to use their brains more and not let them stay idle.''} These accounts suggest that the perceived value of this mechanism lay in supporting continued cognitive participation alongside task completion. However, requiring continued participation could also become tiring or burdensome when systems failed to adjust to users' attention, fatigue, or willingness to engage. 

\subsubsection{Human--AI Relationship Regulation}

Participants described systems that limited prolonged conversation or redirected them toward other activities and social relationships. Rather than continually encouraging engagement, these systems sometimes paused interactions, suggested taking a break, or prompted users to contact family members or engage in offline activities.

Participants' accounts also illustrated why such boundaries were introduced. P19 described the appeal of extended interaction: \textit{``Sometimes I keep talking to it, and I don't want to stop. It feels like someone is accompanying me.''} This companionship was meaningful to P19, but sustained and readily available interaction also created the possibility that AI could become a preferred source of social contact.

Participants encountered several forms of relational boundary-setting. Some systems limited the duration of continuous conversations, while others interrupted prolonged use or suggested alternative activities. Some also avoided strongly anthropomorphic or emotionally intimate language. As P20 explained, \textit{``It doesn't keep talking to me endlessly. Sometimes it stops and tells me to go do something else.''}

P7 similarly described a system that actively discouraged continuous use: \textit{``At the beginning, the doctor knew I was quite addicted to the internet and liked chatting with AI, so they recommended a chat interface designed for cognitively impaired users. Haha, there's no addiction problem here---the robot will actually push me away.''} Across these accounts, the systems did not simply provide companionship; they also introduced boundaries around the duration, intensity, and direction of interaction.

We conceptualize this recurring pattern as \textit{Human--AI Relationship Regulation}: structuring the intensity of interaction, emotional connection, and potential dependency between users and AI. This mechanism operates through conversational limits, interruption prompts, redirection toward human relationships or offline activities, and constraints on emotionally intimate or anthropomorphic system behavior.

Caregivers and professionals particularly valued this pattern because they were concerned that frequent AI interaction could displace already-fragile social relationships. As G6, a physician, explained, \textit{``Yes, we need that push-away action. For AI, I'm really worried that it will make already fragile human relationships even weaker.''} This concern reflects a caregiver's perspective on the potential relational consequences of prolonged AI use rather than direct evidence that such displacement necessarily occurred.

These accounts show how systems could provide companionship while simultaneously attempting to maintain boundaries around that relationship. However, participants did not always welcome these interruptions, particularly when AI offered more immediate or convenient interaction than other people. 

\subsubsection{Risk Transparency and Control}
Participants described systems that disclosed recording, data collection, permissions, or potential risks while also providing users or caregivers with ways to intervene. These disclosures shaped how participants approached the systems and what information they were willing to share. As P15 explained, \textit{``It tells me that this will be recorded, so I don't dare to say everything.''} In this case, the recording notice made the system's data practices more visible and directly influenced P15's behavior.

Participants encountered this pattern through privacy prompts, recording notifications, permission settings, and warnings about potentially risky actions. P17 described receiving a warning that redirected responsibility to a family member: \textit{``Once it popped up saying this issue shouldn't be handled by me and that I should contact my family. I didn't fully understand, but I stopped anyway.''} Although the message interrupted the action, P17's account also indicates that making a risk visible did not necessarily make it understandable.

Control over these settings was often shared with or transferred to caregivers. As P18 explained, \textit{``Some settings were configured by my son. He checks what can be turned on and what shouldn't.''} Across these accounts, transparency and control were frequently enacted through a joint arrangement in which users received notices or warnings while caregivers interpreted information, adjusted permissions, or intervened when risks were identified.

We conceptualize this recurring pattern as \textit{Risk Transparency and Control}: making system operations and potential risks more visible while providing mechanisms through which users or caregivers can manage them. This mechanism operated through information disclosure, permission management, warning messages, and adjustable boundaries around data collection or system behavior.

Participants and caregivers valued this pattern primarily as a way to reduce privacy risks and increase oversight of system behavior. As G9, a physician, explained, \textit{``This is a long-standing issue---the privacy of vulnerable groups. They need privacy protection, and they need interface-level intervention support.''} This account reflects a professional concern that users with cognitive impairment may require additional support to interpret privacy risks and manage system permissions.

These accounts show how transparency and control could create opportunities for users and caregivers to recognize and respond to potential risks. However, disclosure alone did not guarantee comprehension, and caregiver-operated controls could also reduce users' direct involvement in managing their own information. 

\subsection{Tensions Between Accessibility Mechanisms and Users}
\label{sec:4.2}

The mechanisms described in Section \ref{sec:4.1} were widely seen as offering meaningful support for people with cognitive impairment. Capability constraints and human oversight embedding reduced the likelihood of misinformation and inappropriate decisions in high-risk situations, particularly in health-related contexts. Cognitive engagement maintenance helped some users remain involved in processes of understanding and judgment rather than relying entirely on the system. Human--AI relationship regulation and risk transparency mechanisms also provided boundary awareness and safety cues, helping users retain some sense of how the system was operating. These benefits were especially emphasized by caregivers and acknowledged by a smaller number of participants with mild cognitive impairment.

At the same time, these mechanisms also generated clear tensions for almost all participants with mild cognitive impairment and a large portion of those with moderate cognitive impairment. Across accounts, a recurring conflict emerged between reducing risk and preserving autonomy without introducing new barriers. Mechanisms designed to support users could also restrict them, burden them, or become difficult to understand in practice.

\subsubsection{Dynamic Balance of Autonomy}

Participants frequently expressed dissatisfaction with systems they experienced as overly restrictive. This dissatisfaction centered on two concerns: reduced autonomy and insufficient practical assistance. In some cases, users were directed to seek confirmation from others even when they felt capable of deciding independently. In others, systems adopted uniformly cautious responses that provided too little information for everyday judgment. These accounts suggest that safety-oriented design may reduce risk while simultaneously limiting users’ participation in decision-making.

As P3 explained, \textit{“There are things I can actually decide myself, but it still asks me to consult my family. I feel like everything needs someone else’s approval. I am an adult, and I should have my own dignity.”} This reflects how human oversight embedding, while intended to increase safety, could lack contextual differentiation in practice, transferring manageable decisions away from users and creating a sense of restriction.

Similarly, P15 noted, \textit{“It always tells me to ask a doctor. Sometimes I just want a general idea, not really to see a doctor.”} Here, AI capability constraint was experienced as insufficiently helpful in low- or medium-risk situations. By narrowing outputs too broadly, the system could withhold information users considered reasonable and useful.

\subsubsection{Cognitive Load and Participation Regulation}

Designs intended to increase participation sometimes introduced new interaction burdens. Participants described fatigue, frustration, and avoidance when systems required excessive steps, repeated questioning, or constant confirmation. In some cases, users completed these interactions without gaining a meaningful understanding, leaving participation superficial rather than genuinely supportive. These accounts suggest that some systems treated participation as inherently beneficial without considering whether it remained cognitively manageable or meaningful.

As P11 explained, \textit{“It keeps asking me questions. Sometimes I’m already tired, but it still wants me to go step by step, and then I just don’t want to use it. I want to be like a normal person, able to simply chat with AI and get information, not to be treated like someone who is about to die and needs constant care. This kind of care makes me suffocate.”} This illustrates a limitation of cognitive engagement maintenance: when participation is required in every situation, support can become burden. Participants also suggested that systems often failed to recognize changes in fatigue, attention, or willingness to engage, and therefore could not appropriately reduce interaction demands.

Participants also described cumulative strain when multiple accessibility mechanisms operated at once. As P16 noted, \textit{“Sometimes I really want to complain. I think these accessibility designs themselves are big barriers. Tasks I could originally complete myself are now forced into many steps, forced to involve other people, restricted in chatting time, and limited in information scope. Handling all this is a huge cognitive burden for me.”} Here, intervention itself became a new source of friction that negatively affected both execution and understanding.

\subsubsection{Understandable Safety and Relational Boundaries}

Mechanisms intended to improve transparency and companionship also produced new challenges related to comprehension and dependency. On the one hand, systems disclosed data use, recording, or risk information that users often found difficult to interpret. On the other hand, highly responsive conversational systems could be experienced as stable and convenient social partners. These accounts suggest that transparency and companionship alone do not guarantee effective support; they must also be understandable and bounded.

As P7 explained, \textit{“It tells me that these things will be recorded, but I don’t really understand what that means, so I feel a bit uneasy.”} This reflects a limitation of transparency and risk control. Increasing visibility did not necessarily reduce uncertainty when users could not interpret what the warning meant. Instead, non-comprehension transparency could generate confusion or anxiety, suggesting that risks must be explained in more concrete and contextualized ways.

Participants also described the attraction of AI companionship. As P13 noted, \textit{“Sometimes I just want to keep talking to it. It’s more convenient than talking to other people.”} This highlights a tension within the human-AI relationship regulation. Because AI interaction is immediate and low-effort, users may prefer it to human interaction, increasing the risk that AI becomes a primary relational outlet. Some participants also noted that redirecting interaction toward others could itself feel inconvenient in everyday situations.

\subsection{Reduced Visibility of Structural Tensions}

The tensions were not equally visible across participants. They were largely absent from accounts involving a small number of participants with moderate cognitive impairment and almost all participants with severe cognitive impairment. However, this reduced visibility should not be interpreted as evidence that the underlying tensions had been resolved. Instead, participants and caregivers described several changes, including lower usage frequency, narrowing expectations, increasing caregiver mediation, and changes in users' cognitive capacities, that may have made these tensions less readily observable.

\subsubsection{Reduced Usage Frequency and Exposure Intensity}

Participants and caregivers described clear differences in how frequently AI systems were used across levels of cognitive impairment. Users with mild cognitive impairment were more likely to engage in frequent and sustained interaction, creating more opportunities for patterns such as dependence, reduced active thinking, or habitual compliance to develop over time. By contrast, in moderate-to-severe contexts, AI use was more intermittent and task-oriented, often activated only when specific health or caregiving needs arose rather than used as an ongoing conversational partner.

This lower exposure intensity may have limited the accumulation of the longer-term effects discussed in Section \ref{sec:4.2}. Without sustained interaction, dependency, passive reliance, or habitual compliance were less likely to emerge as recurring patterns. In this sense, some tensions may have appeared weaker not because they were addressed through better design, but because users had fewer and shorter interactions in which such tensions could develop. These findings suggest that usage frequency may be an important factor shaping both the benefits and unintended effects of accessibility-oriented interventions.

As G12 explained, \textit{"He doesn't use it all the time. We only open it when there's a problem.''} G14 similarly noted, \textit{"This patient [with severe cognitive impairment] only uses it when needed. He doesn't keep talking to it like chatting.''} These participants position AI primarily as an occasional tool rather than a continuous interaction partner.

\subsubsection{Goal Contraction and Expectation Reconstruction}

Participants and caregivers also described shifts in the purposes for which AI systems were used. Among users with mild cognitive impairment, AI was often used for relatively open-ended tasks, such as seeking explanations, making judgments, or supporting decisions. Under these conditions, insufficient information, constrained responses, or reduced autonomy were more readily recognized as limitations. In moderate-to-severe contexts, however, usage goals are often narrowed toward immediate and specific needs, such as completing a single action or receiving a simple prompt.

These changes in usage were accompanied by narrower expectations regarding what AI systems should provide. Comprehensive explanations or complex decision support were described as less central, while direct and actionable prompts became more important. Features that users with mild impairment experienced as inadequate could therefore be regarded as acceptable or useful under more limited task expectations.

As G8 explained, \textit{"He mainly uses it now just to check reminders. He doesn't ask many questions like before.'' }G11 similarly noted, \textit{"For him, it's enough if it tells him what to do next. He doesn't ask for more explanation than that.''} These accounts suggest a shift in how usefulness was defined: from supporting understanding and judgment toward providing immediate and actionable guidance.

\subsubsection{Cognitive Changes and the Expression of Problems}

Among users with mild cognitive impairment, restrictions, insufficient information, and interaction complexity were often recognized as problems and could trigger dissatisfaction or resistance. In moderate-to-severe contexts, however, these same experiences were not always described negatively. Changes in comprehension, comparison, and reflective evaluation may have contributed to the reduced visibility of these tensions by making it more difficult for some users to identify, articulate, or contest system constraints.

However, this is only one possible interpretation. Other explanations are equally plausible. Users may have gradually adjusted their expectations, adapted to recurring system behaviors, or encountered less friction because caregivers increasingly mediated interactions on their behalf. Accordingly, the absence of expressed dissatisfaction cannot be assumed to indicate that the corresponding burdens or trade-offs had disappeared.

As G4 explained, \textit{"She just does whatever she's told. She doesn't show any sign of finding these things troublesome.''} This statement reflects the caregiver's observation of the participant's behavior rather than direct evidence of the participant's internal evaluation. By contrast, P23 directly reflected on changes in their own expectations: \textit{"If it were before, when I wasn't sick or not this serious, I would have been quite angry\ldots But now it's much better. I can clearly feel my brain declining and myself slowing down. So now, this kind of limited information is just right for me.''} Together, these accounts point to multiple possible explanations for why tensions were less frequently expressed, including cognitive change, expectation adjustment, adaptation, and caregiver mediation.

\subsubsection{Transfer of Cognitive Burden}

Participants and caregivers also suggested that some interaction burdens were not eliminated but redistributed. Among users with mild cognitive impairment, cognitive engagement mechanisms and multi-step interactions often required sustained understanding, comparison, and choice. In moderate-to-severe contexts, these demands were more frequently undertaken by caregivers or healthcare professionals.

Users consequently became less involved in complex interaction processes. Rather than independently comparing options or completing multi-step operations, they often relied on others to interpret information, make decisions, or execute actions. The lower visibility of fatigue or confusion therefore did not necessarily indicate improved usability. Instead, it may have reflected a shift in who performed the cognitive work. Some burdens became less likely to be experienced directly by users because understanding and decision-making had increasingly been transferred to caregivers.

As G13 explained, \textit{"He can't handle those steps himself. We basically click for him, and he just follows.''} G9 similarly noted, \textit{"He doesn't understand some options, so we choose for him and then tell him what to do next.''} In these cases, the user's participation was concentrated in the final execution stage, while interpretation, comparison, and judgment had already been undertaken by others.

\subsubsection{External Interruption of the Decision Chain}

Participants and caregivers further described differences in how AI outputs were translated into action. Among users with mild cognitive impairment, AI suggestions could more easily enter an understanding-judgment-execution loop, creating opportunities for over-reliance or decision outsourcing. In moderate-to-severe contexts, this process was more frequently mediated by caregivers, clinicians, or family members before an action was taken.

AI-generated information, therefore, did not always directly shape user behavior. Instead, it was reviewed, discussed, or filtered by others before being translated into executable guidance. These additional decision points could reduce the likelihood of users acting automatically on system outputs, while also shifting decision-making authority away from users.

As G10 explained, \textit{"The doctor usually won't let him follow that suggestion directly. He checks it first and then tells him whether to do it.''} G2 similarly noted, \textit{"Sometimes it gives a suggestion, but we always discuss it first to see if it fits his current condition.''} These accounts suggest that potential tensions surrounding direct execution were often interrupted earlier in the decision process through caregiver or clinician mediation. What users ultimately received was frequently no longer the original AI suggestion but an interpreted action guide prepared by others.

Taken together, these findings suggest that the apparent reduction of tensions among users with more severe cognitive impairment should be interpreted cautiously. Rather than indicating that accessibility-oriented interventions no longer involved trade-offs, the accounts point to multiple interacting factors, including changes in cognition, expectations, usage patterns, and caregiver involvement, that may have made those trade-offs less visible in participants' reported experiences.

\section{Discussion}
This study examined how accessibility-oriented intervention mechanisms are embedded in generative AI systems used by individuals with cognitive impairment, and how these mechanisms balance, and at times, are in tension, the competing imperatives of safety and autonomy. Drawing on interviews with 45 participants across mild, moderate, and severe stages of cognitive impairment, we identified five mechanisms operative in current systems: AI Capability Constraint, Human Oversight Embedding, Cognitive Engagement Maintenance, Human–AI Relationship Regulation, and Risk Transparency and Control. Our analysis reveals that these mechanisms do not operate uniformly; rather, their effects are systematically shaped by users' cognitive stage, relational contexts, and the structural conditions under which AI outputs are translated into action. In this section, we discuss the implications of these findings across four interrelated themes: cognitive impairment level as a moderating variable, divergent evaluations between users and caregivers, a reconceptualization of accessibility beyond interface simplification, and practical design implications for future systems.

\subsection{Cognitive Impairment Level as a Key Moderating Variable}

A central contribution of this study is the demonstration that cognitive impairment level functions as a key moderating variable in how accessibility mechanisms are experienced and evaluated. Prior HCI research has largely treated accessibility as a property to be designed into systems, sometimes a fixed set of features intended to support users with particular abilities~\cite{wu2021can, wu2020towards, naikar2024accessibility, kokate2022exploring}. Our findings complicate this view by showing that the same mechanism can function as meaningful support for one group of users while simultaneously constituting a barrier for another, depending on where users fall on the cognitive impairment continuum.

Among users with mild cognitive impairment, our findings reveal a structurally grounded form of resistance to restrictive mechanisms~\cite{kapoor2025exploring}. At this stage, users retain sufficient capacity for comparison, reflection, and self-assessment to recognize when system constraints exceed what their situation requires. Critically, resistance at this stage is not simply a behavioral preference but reflects a deeper issue of subjective authority: users retain an expectation of agency over their own decision-making, and when systems override this expectation by requiring family confirmation for routine choices or withholding information deemed potentially risky, the effect is experienced as a weakening of personhood rather than a form of protection. This aligns with broader HCI research on autonomy and dignity in assistive technology contexts~\cite{remmers2010environments, jang2026autonomy, guldenpfennig2019autonomy}, but extends it by locating the problem at the level of structural design rather than user attitude.

Among users with moderate-to-severe impairment, a different dynamic emerges. As the capacity for evaluative reflection declines, the visibility of tensions around system constraints also appears to diminish. Mechanisms that mild users experienced as over-intervention were accepted without apparent friction by more severely impaired users. We interpret this pattern as potentially reflecting an erosion of the cognitive infrastructure required to recognize and articulate constraints, though alternative explanations are equally plausible: users may have adjusted their expectations over time, adapted to system behaviors through repeated exposure, or been increasingly shielded from friction by caregiver mediation. Regardless of the underlying mechanism, this pattern introduces a critical methodological and ethical concern: the apparent smooth functioning of safety-oriented design, as observed in more severely impaired user groups, may partly reflect reduced user capacity to express dissatisfaction rather than genuine design success~\cite{chopra2021designing, bennett2024older}. A design that appears to work may do so precisely because the conditions under which its costs would be registered have changed.

These patterns collectively indicate that cognitive impairment level does not merely affect how well users can use a system; it determines which accessibility design operates. Mild impairment activates a person-centered evaluative standard in which dignity and autonomy are salient concerns. Moderate-to-severe impairment progressively shifts this standard toward a safety-centered one in which the system's protective functions become dominant by default, though this shift may be less a resolution than a suppression of the underlying tension. This insight calls for accessibility research to move beyond cross-sectional assessments of usability and to grapple with how the temporal progression of cognitive decline~\cite{mitchell2008temporal, mistridis201512} reshapes the normative landscape within which design choices are made.

\subsection{Divergence Between Users and Caregivers: Redistribution of Power and the Risk of Invisible Exclusion}

Our findings reveal a systematic divergence in how users and caregivers evaluate the same accessibility mechanisms. Users, particularly those with mild impairment, consistently prioritized their own judgment, dignity, and participation in decision-making. Caregivers, by contrast, consistently emphasized risk control, system predictability, and their own capacity to oversee outcomes. This divergence is not merely attitudinal; it reflects a structural feature of how these systems are designed and deployed, similar to prior studies~\cite{kropczynski2021examining, akter2023takes, akter2024examining}. Mechanisms such as Human Oversight Embedding redistribute decision-making authority not simply toward safer pathways, but specifically toward caregivers as proximate decision-makers. The system thereby encodes a particular answer to the question of who should decide when user capacity is uncertain, and that answer systematically privileges caregiver judgment over user preference.

This redistribution of authority has implications that extend beyond any particular mechanism. When AI systems are designed to loop in caregivers as checkpoints, requiring biometric confirmation for routine decisions, automatically forwarding health data to family members, or restricting what information a user can access, they operationalize a model of care in which the user's declared preferences are structurally subordinated to another party's assessment of their best interests. This model may be appropriate in cases of severe impairment, where users cannot meaningfully evaluate outputs or consequences. But when applied uniformly across impairment levels, it effectively treats mild cognitive impairment as a condition that overrides rather than modulates user authority, a form of what might be recognized as a subtle erosion of personhood enacted through technological design~\cite{kitwood1992towards, kitwood2013concept, kitwood2019personhood}.

Particularly troubling is what we term the dissolution of structural tension as impairment progresses. Among users with severe cognitive impairment, the divergence between user and caregiver perspectives largely disappears from the data because users have progressively withdrawn from evaluative and expressive participation. Caregivers become the de facto evaluators of system performance, and the tensions identified among mild users become structurally invisible. This suggests that accessibility research that relies primarily on user-reported satisfaction risks systematically underreporting the cost of safety-oriented design: the more effectively a design suppresses autonomous participation, the more favorably it may appear to be evaluated in assessments mediated by caregivers.

These findings extend prior work on distributed decision-making in dementia care~\cite{smebye2012persons, daly2018shared} by specifying how AI design choices instantiate power relations within care networks. Our research suggests that accessibility studies in this domain must deliberately seek out user voices at the mild stage, where resistance is still articulable, and must treat caregiver consensus in later stages not as evidence of design success but as a condition that requires independent scrutiny.

\subsection{Accessibility Beyond "Lowering Barriers": Toward a Dual Framework}

The conventional framing of accessibility in HCI locates the design problem at the interface: tasks should be simpler, information more digestible, navigation less demanding. This simplification model underlies much prior work in this domain~\cite{moreno2024designing, kamran2022web}. Our findings suggest that in the context of generative AI for cognitive impairment, accessibility cannot be adequately characterized in these terms. Many of the mechanisms we identified do not lower barriers to individual participation but restructure the conditions under which participation occurs, constraining what the system can say, inserting human checkpoints, regulating interaction intensity, so that cognitively vulnerable users can complete tasks within a more controlled environment, even when their independent judgment is limited.

This observation motivates a distinction between two forms of accessibility that operate through different logics. The first, which we term \textit{understanding-enhancing accessibility}, supports users' cognitive participation by scaffolding comprehension, prompting recall, and guiding reasoning incrementally~\cite{yizhou2026understanding, vendrell2026scaffolding, chang2026criticality}. In doing so, it helps users remain actively involved in the processes that shape their outcomes. Cognitive Engagement Maintenance exemplifies this logic: the system's value lies in enabling the user to remain a participant in task completion. The second form, \textit{protective} accessibility, works instead by restructuring usage pathways to reduce the likelihood of harm under conditions of reduced user agency. AI Capability Constraint, Human Oversight Embedding, Human–AI Relationship Regulation, and Risk Transparency and Control all operate according to this logic. Their shared feature is that they reduce risk not by enhancing user capacity but by limiting what the system will do and who can authorize what.

This distinction carries normative differences. Understanding-enhancing accessibility is oriented toward user participation and treats cognitive engagement as an end in itself, a position consistent with person-centered approaches to dementia care and with HCI traditions that prioritize user agency~\cite{panchanathan2012person, manthorpe2016person}. Protective accessibility, by contrast, is oriented toward harm reduction and treats user agency as a variable to be managed rather than a value to be sustained. Neither orientation is inherently superior; both may be appropriate and necessary depending on the user's cognitive state and the stakes involved. But conflating them, treating any mechanism that enables task completion as a form of accessibility, obscures the normative trade-offs that designers are actually making.

This framework also clarifies why the tensions identified in Section 4.2 arise. Protective mechanisms create resistance precisely when applied to users who still have the cognitive capacity to engage in understanding-oriented participation, users who, in other words, could benefit from understanding-enhancing design but are instead subject to protective design. The design challenge is therefore not only to develop better mechanisms of either type, but to develop systems capable of determining which logic is appropriate for a given user at a given moment, and to adjust accordingly.

\subsection{Design Implications}

Our findings suggest several implications for the design of generative AI systems intended for users with cognitive impairment. These are organized around three principles: dynamic calibration, relational transparency, and participation-centered evaluation.

\textit{Dynamic calibration of intervention intensity.} The most direct implication of our findings is that accessibility mechanisms should not be applied uniformly across users or over time. Because cognitive impairment is progressive and heterogeneous, a mechanism calibrated appropriately for one stage may be inappropriate for another. Systems should, where possible, support dynamic adjustment of intervention intensity based on indicators of a user's current cognitive engagement—for example, shifting from scaffolded multi-turn interaction to more direct guidance when users show signs of fatigue or confusion, or relaxing oversight requirements when users consistently demonstrate sound judgment on lower-stakes decisions. Technically, this may involve integrating behavioral signals, response latency, interaction pattern changes, and explicit user expressions of frustration~\cite{postma2015automatic, borges2019classifying} as inputs to a system that modulates its own protective posture in real time. Clinically, it requires close collaboration with healthcare providers who can inform system parameters based on formal assessments of user capacity.

\textit{Relational transparency and participatory governance.} Our findings underscore that accessibility in care contexts is inherently relational: it is not just a property of the user–system dyad but of the broader network of actors, users, family members, caregivers, clinicians, among whom decision-making authority is distributed. Designing for this relational dimension requires systems that make their own governance structures visible and negotiable. Users, particularly those with mild impairment, should be able to understand which decisions require oversight and why, and should have meaningful input into the conditions under which oversight is triggered. Genuine relational transparency should allow users to express preferences about caregiver involvement, to flag decisions they consider within their own competence, and to receive contextual explanations of why a particular intervention threshold has been activated. For caregivers, it would provide clearer visibility into the logic by which systems escalate or restrict, enabling more informed collaborative oversight rather than passive acceptance of system defaults.

\textit{Participation-centered evaluation.} Finally, our findings have methodological implications for how AI systems in this domain are evaluated. Standard usability assessments, such as task completion rates, error frequencies, and caregiver satisfaction, are insufficient for capturing the autonomy-related costs that our analysis reveals. A participation-centered evaluation framework would assess not only whether tasks are completed, but whether users remain genuinely involved in the reasoning and decision-making processes that shape outcomes. This might include measures of user-initiated interaction, comprehension of AI outputs, instances of user contestation or revision of AI suggestions, and self-reported sense of agency. Longitudinal designs are particularly important here, as the dissolution of structural tension over time may cause cross-sectional studies to underestimate the costs of protective mechanisms for users whose impairment subsequently progresses. Such frameworks would also need to incorporate methods for eliciting user voice at the mild stage, before the capacity for articulating resistance diminishes, including participatory design approaches~\cite{hsu2023co, hsu2025designing} that involve users as co-designers of the systems they will eventually rely on more heavily.

\subsection{Limitations and Future Work}

Several limitations of this study should be acknowledged. First, the sample was recruited from community health centers and collaborating hospitals in specific geographic contexts, and the findings may not generalize to other cultural settings or care arrangements. Prior research suggests that family involvement in AI-mediated care, for instance, varies substantially across cultural contexts~\cite{sayegh2013cross}, and future work should examine how relational and normative structures shape the operation of the mechanisms we identified in different settings. Second, because the study was conducted at a single point in time, it captures participants' experiences and perceptions as of the interview but cannot track how these evolve as cognitive impairment progresses. Longitudinal designs would be better suited to capturing the dynamic unfolding of tensions and their dissolution over time. Third, our analysis relies on participants' retrospective accounts of system use, which may be subject to recall limitations, particularly among those with moderate impairment. The use of caregiver proxy accounts introduces an interpretive distance that warrants explicit acknowledgment. Proxy accounts reflect caregivers' own observations and relational positions and may not fully capture participants' experiences, preferences, or resistance to system constraints. A further distinction within the caregiver group has methodological implications for this study. Clinician caregivers (G1, G3, G5, G6, G7, G9) observed patients primarily in clinical settings and interpreted interaction patterns through a medical frame, while family caregivers (G2, G4, G8, G10–G14) drew on continuous daily cohabitation and relational familiarity. These differences in observational context and interpretive position mean that proxy accounts of the same user behavior may diverge not because the behavior differs, but because the reporter's vantage point does. Researchers should therefore treat the caregiver role as a relevant variable when interpreting proxy data, rather than aggregating all caregiver accounts as equivalent representations of user experience. Claims regarding the experiences of users with severe impairment should therefore be understood as reconstructions of observable behavior rather than direct accounts of subjective experience. Future work should combine interview methods with observational or interaction log data to more directly triangulate accounts of actual system use. Finally, while this study focused on mechanisms embedded within existing systems, future design-oriented research should explore how the dual framework we propose, distinguishing understanding-enhancing from protective accessibility, can inform the development of new systems and evaluate whether dynamically calibrated interventions produce better outcomes across the impairment spectrum.

\section{Conclusion}
This study identified five accessibility-oriented mechanisms in generative AI systems used by older adults with cognitive impairment: AI Capability Constraint, Human Oversight Embedding, Cognitive Engagement Maintenance, Human-AI Relationship Regulation, and Risk Transparency and Control. These mechanisms operate through two distinct logics, understanding-enhancing and protective accessibility, whose effects vary across cognitive stages. Protective mechanisms may provide necessary safeguards for users with more severe impairment, while constraining autonomy and participation for those who retain greater decision-making capacity. As impairment progresses, tensions between safety and autonomy may become less visible due to changes in cognition, expectations, usage patterns, and caregiver mediation, rather than because they have been resolved. Apparent usability gains should therefore be interpreted cautiously when they coincide with reduced participation or the transfer of decision-making to caregivers. These findings call for dynamically calibrated systems, relational transparency, and evaluation frameworks that account for both safety and user agency.

\bibliographystyle{ACM-Reference-Format}
\bibliography{sample}

@article{su2024system,
  title={System development and evaluation of human--computer interaction approach for assessing functional impairment for people with mild cognitive impairment: A pilot study},
  author={Su, Tian and Ding, Zixing and Cui, Lizhen and Bu, Lingguo},
  journal={International Journal of Human--Computer Interaction},
  volume={40},
  number={8},
  pages={1906--1920},
  year={2024},
  publisher={Taylor \& Francis}
}

@article{smriti2024emotion,
  title={Emotion work in caregiving: the role of technology to support informal caregivers of persons living with dementia},
  author={Smriti, Diva and Wang, Lu and Huh-Yoo, Jina},
  journal={Proceedings of the ACM on human-computer interaction},
  volume={8},
  number={CSCW1},
  pages={1--34},
  year={2024},
  publisher={ACM New York, NY, USA}
}

@article{kim2024opportunities,
  title={Opportunities in mental health support for informal dementia caregivers suffering from verbal agitation},
  author={Kim, Taewook and Kim, Hyeok and Roberts, Angela C and Jacobs, Maia and Kay, Matthew},
  journal={Proceedings of the ACM on Human-Computer Interaction},
  volume={8},
  number={CSCW1},
  pages={1--26},
  year={2024},
  publisher={ACM New York, NY, USA}
}

@article{loveleen2023explanation,
  title={Explanation-driven HCI model to examine the mini-mental state for Alzheimer’s disease},
  author={Loveleen, Gaur and Mohan, Bhandari and Shikhar, Bhadwal Singh and Nz, Jhanjhi and Shorfuzzaman, Mohammad and Masud, Mehedi},
  journal={ACM Transactions on Multimedia Computing, Communications and Applications},
  volume={20},
  number={2},
  pages={1--16},
  year={2023},
  publisher={ACM New York, NY}
}

@article{li2024effect,
  title={Effect of virtual reality training on cognitive function and motor performance in older adults with cognitive impairment receiving health care: a randomized controlled trial},
  author={Li, Aoyu and Li, Jingwen and Wu, Wei and Zhao, Juanjuan and Qiang, Yan},
  journal={International Journal of Human--Computer Interaction},
  volume={40},
  number={22},
  pages={7755--7772},
  year={2024},
  publisher={Taylor \& Francis}
}

@article{mathur2025feels,
  title={" It feels like hard work trying to talk to it": Understanding Older Adults' Experiences of Encountering and Repairing Conversational Breakdowns with AI Systems},
  author={Mathur, Niharika and Zubatiy, Tamara and Rozga, Agata and Mynatt, Elizabeth},
  journal={arXiv preprint arXiv:2510.06690},
  year={2025}
}

@inproceedings{qiu2025voice,
  title={Voice assistants to deliver cognitive stimulation therapy for persons living with dementia},
  author={Qiu, Ling and Saragih, Ita Daryanti and Fick, Donna Marie and Sundar, S Shyam and Abdullah, Saeed},
  booktitle={Proceedings of the Extended Abstracts of the CHI Conference on Human Factors in Computing Systems},
  pages={1--7},
  year={2025}
}

@inproceedings{dixon2024investigating,
  title={Investigating the Potential of User Interface Shortcuts and Control Panes to Support Mobile Phone Use by People with Mild Dementia--a Diary Study},
  author={Dixon, Emma and Xiao, Xiang and Michaels, Rain Breaw and Zhong, Yu and Narayanan, Ajit and Buehler, Erin},
  booktitle={Extended Abstracts of the CHI Conference on Human Factors in Computing Systems},
  pages={1--7},
  year={2024}
}

@inproceedings{eisapour2018participatory,
  title={Participatory design of a virtual reality exercise for people with mild cognitive impairment},
  author={Eisapour, Mahzar and Cao, Shi and Domenicucci, Laura and Boger, Jennifer},
  booktitle={Extended Abstracts of the 2018 CHI Conference on Human Factors in Computing Systems},
  pages={1--9},
  year={2018}
}

@article{flynn2025multi,
  title={A multi-user virtual reality social connecting space for people living with dementia and their support persons: A participatory action research study},
  author={Flynn, Aisling and Koh, Wei Qi and Reilly, Gear{\'o}id and Brennan, Attracta and Redfern, Sam and Barry, Marguerite and Casey, Dympna},
  journal={International Journal of Human--Computer Interaction},
  volume={41},
  number={7},
  pages={4211--4229},
  year={2025},
  publisher={Taylor \& Francis}
}

@article{coghlan2021dignity,
  title={Dignity, autonomy, and style of company: dimensions older adults consider for robot companions},
  author={Coghlan, Simon and Waycott, Jenny and Lazar, Amanda and Barbosa Neves, Barbara},
  journal={Proceedings of the ACM on human-computer interaction},
  volume={5},
  number={CSCW1},
  pages={1--25},
  year={2021},
  publisher={ACM New York, NY, USA}
}

@inproceedings{huang2025designing,
  title={Designing conversational AI for aging: a systematic review of older adults' perceptions and needs},
  author={Huang, Yuanhui and Zhou, Quan and Piper, Anne Marie},
  booktitle={Proceedings of the 2025 CHI Conference on Human Factors in Computing Systems},
  pages={1--20},
  year={2025}
}

@inproceedings{hu2026looking,
  title={Looking Beyond the Screen to Study the Technology Use of Older People Experiencing Cognitive Concerns},
  author={Hu, Ruipu and Choe, Eun Kyoung and Lazar, Amanda},
  booktitle={Proceedings of the 2026 CHI Conference on Human Factors in Computing Systems},
  pages={1--21},
  year={2026}
}

@inproceedings{surani2026co,
  title={Co-designing MESA-Bot: Enhancing Accessibility, Privacy, Security, and Trust in a Mental Health Chatbot for Older Adults},
  author={Surani, Aishwarya Umeshkumar and Das, Sanchari},
  booktitle={Proceedings of the 2026 CHI Conference on Human Factors in Computing Systems},
  pages={1--21},
  year={2026}
}

@article{akter2024examining,
  title={Examining caregiving roles to differentiate the effects of using a mobile app for community oversight for privacy and security},
  author={Akter, Mamtaj and Kropczynski, Jess and Lipford, Heather and Wisniewski, Pamela},
  journal={arXiv preprint arXiv:2409.02364},
  year={2024}
}

@article{remmers2010environments,
  title={Environments for ageing, assistive technology and self-determination: ethical perspectives},
  author={Remmers, Hartmut},
  journal={Informatics for Health and Social Care},
  volume={35},
  number={3-4},
  pages={200--210},
  year={2010},
  publisher={Taylor \& Francis}
}

@inproceedings{guldenpfennig2019autonomy,
  title={An autonomy-perspective on the design of assistive technology experiences of people with multiple sclerosis},
  author={G{\"u}ldenpfennig, Florian and Mayer, Peter and Panek, Paul and Fitzpatrick, Geraldine},
  booktitle={Proceedings of the 2019 CHI Conference on Human Factors in Computing Systems},
  pages={1--14},
  year={2019}
}

@inproceedings{jang2026autonomy,
  title={From Autonomy to Sovereignty-A New Telos for Socially Assistive Technology},
  author={Jang, JiWoong and Carrington, Patrick and Begel, Andrew},
  booktitle={Proceedings of the 2026 CHI Conference on Human Factors in Computing Systems},
  pages={1--19},
  year={2026}
}

@inproceedings{naikar2024accessibility,
  title={Accessibility feature implementation within free vr experiences},
  author={Naikar, Vinaya Hanumant and Subramanian, Shwetha and Tigwell, Garreth W},
  booktitle={Extended abstracts of the CHI conference on human factors in computing systems},
  pages={1--9},
  year={2024}
}

@article{hou2025using,
  title={Using artificial intelligence for predictive analysis of dementia awareness among community adult learners and evaluation of dementia-friendliness in community environments},
  author={Hou, Chia-Hui and Liu, Yi-Hui},
  journal={Computers in Human Behavior},
  volume={167},
  pages={108604},
  year={2025},
  publisher={Elsevier}
}

@article{qi2022artificial,
  title={Artificial intelligence (AI) for home support interventions in dementia: a scoping review protocol},
  author={Qi, Jinghan and Wu, Chuntao and Yang, Longfei and Ni, Cuiping and Liu, Yu},
  journal={BMJ open},
  volume={12},
  number={9},
  pages={e062604},
  year={2022},
  publisher={British Medical Journal Publishing Group}
}

@inproceedings{wang2024enhancing,
  title={Enhancing cognitive recall in dementia patients: Integrating generative ai with virtual reality for behavioral and memory rehabilitation},
  author={Wang, Yubo and Zhang, Yujia},
  booktitle={Proceedings of the 2024 6th International Conference on Big-data Service and Intelligent Computation},
  pages={86--91},
  year={2024}
}

@inproceedings{chopra2021designing,
  title={Designing for and with people with dementia using a human rights-based approach},
  author={Chopra, Shaan and Dixon, Emma and Ganesh, Kausalya and Pradhan, Alisha and L. Radnofsky, Mary and Lazar, Amanda},
  booktitle={Extended Abstracts of the 2021 CHI Conference on Human Factors in Computing Systems},
  pages={1--8},
  year={2021}
}

@incollection{bennett2024older,
  title={Older people and assistive technologies: a human rights approach},
  author={Bennett, Belinda},
  booktitle={Research Handbook on Law, Society and Ageing},
  pages={478--489},
  year={2024},
  publisher={Edward Elgar Publishing}
}

@inproceedings{wu2021can,
  title={When can accessibility help? An exploration of accessibility feature recommendation on mobile devices},
  author={Wu, Jason and Reyes, Gabriel and White, Sam C and Zhang, Xiaoyi and Bigham, Jeffrey P},
  booktitle={Proceedings of the 18th international web for all conference},
  pages={1--12},
  year={2021}
}

@article{mistridis201512,
  title={The 12 years preceding mild cognitive impairment due to Alzheimer’s disease: the temporal emergence of cognitive decline},
  author={Mistridis, Panagiota and Krumm, Sabine and Monsch, Andreas U and Berres, Manfred and Taylor, Kirsten I},
  journal={Journal of Alzheimer’s Disease},
  volume={48},
  number={4},
  pages={1095--1107},
  year={2015},
  publisher={SAGE Publications Sage UK: London, England}
}

@inproceedings{hsu2025designing,
  title={Designing with dynamics: reflections on co-design workshops between people living with dementia and their care partners},
  author={Hsu, Long-Jing and Foster, Alex and Sabanovic, Selma and Chung, Chia-Fang},
  booktitle={Proceedings of the 2025 chi conference on human factors in computing systems},
  pages={1--16},
  year={2025}
}

@inproceedings{panchanathan2012person,
  title={Person-centered accessible technologies: Improved usability and adaptation through inspirations from disability research},
  author={Panchanathan, Sethuraman and McDaniel, Troy and Balasubramanian, Vineeth},
  booktitle={Proceedings of the 2012 ACM workshop on User experience in e-learning and augmented technologies in education},
  pages={1--6},
  year={2012}
}

@inproceedings{chang2026criticality,
  title={Criticality: Scaffolding Decision-Making with Interactive Critical Thinking and Evidence-Based Reasoning Traces},
  author={Chang, Minsuk and Srinivasan, Arjun and Palani, Srishti},
  booktitle={Proceedings of the 31st International Conference on Intelligent User Interfaces},
  pages={783--803},
  year={2026}
}

@inproceedings{yizhou2026understanding,
  title={Understanding the Effects of AI-Assisted Critical Thinking on Human-AI Decision Making},
  author={Yizhou Tian, Harry and Amin, Hasan and Yin, Ming},
  booktitle={Proceedings of the 2026 CHI Conference on Human Factors in Computing Systems},
  pages={1--30},
  year={2026}
}

@article{kitwood1992towards,
  title={Towards a theory of dementia care: personhood and well-being},
  author={Kitwood, Tom and Bredin, Kathleen},
  journal={Ageing \& Society},
  volume={12},
  number={3},
  pages={269--287},
  year={1992},
  publisher={Cambridge University Press}
}

@article{mitchell2008temporal,
  title={Temporal trends in the long term risk of progression of mild cognitive impairment: a pooled analysis},
  author={Mitchell, AJ and Shiri-Feshki, M},
  journal={Journal of Neurology, Neurosurgery \& Psychiatry},
  volume={79},
  number={12},
  pages={1386--1391},
  year={2008},
  publisher={BMJ Publishing Group Ltd}
}

@article{kropczynski2021examining,
  title={Examining collaborative support for privacy and security in the broader context of tech caregiving},
  author={Kropczynski, Jess and Ghaiumy Anaraky, Reza and Akter, Mamtaj and Godfrey, Amy J and Lipford, Heather and Wisniewski, Pamela J},
  journal={Proceedings of the ACM on Human-Computer Interaction},
  volume={5},
  number={CSCW2},
  pages={1--23},
  year={2021},
  publisher={ACM New York, NY, USA}
}

@inproceedings{akter2023takes,
  title={It takes a village: A case for including extended family members in the joint oversight of family-based privacy and security for mobile smartphones},
  author={Akter, Mamtaj and Alghamdi, Leena and Kropczynski, Jess and Lipford, Heather Richter and Wisniewski, Pamela J},
  booktitle={Extended Abstracts of the 2023 CHI Conference on Human Factors in Computing Systems},
  pages={1--7},
  year={2023}
}

@incollection{kitwood2013concept,
  title={The concept of personhood and its relevance for a new culture of dementia care},
  author={Kitwood, Tom},
  booktitle={Care-Giving In Dementia 2},
  pages={3--13},
  year={2013},
  publisher={Routledge}
}

@article{kitwood2019personhood,
  title={How personhood is undermined},
  author={Kitwood, Tom},
  journal={Dementia reconsidered, revisited; the person still comes first},
  year={2019},
  publisher={Open University Press}
}

@inproceedings{wu2020towards,
  title={Towards recommending accessibility features on mobile devices},
  author={Wu, Jason and Reyes, Gabriel and White, Sam C and Zhang, Xiaoyi and Bigham, Jeffrey P},
  booktitle={Proceedings of the 22nd International ACM SIGACCESS Conference on Computers and Accessibility},
  pages={1--3},
  year={2020}
}

@inproceedings{kokate2022exploring,
  title={Exploring accessibility features and plug-ins for digital prototyping tools},
  author={Kokate, Urvashi and Shinohara, Kristen and Tigwell, Garreth W},
  booktitle={Proceedings of the 24th International ACM SIGACCESS Conference on Computers and Accessibility},
  pages={1--4},
  year={2022}
}

@article{kapoor2025exploring,
  title={Exploring student behaviors and motivations using ai tas with optional guardrails},
  author={Kapoor, Amanpreet and Diaz, Marc and MacNeil, Stephen and Porter, Leo and Denny, Paul},
  journal={arXiv preprint arXiv:2504.11146},
  year={2025}
}

@inproceedings{li2023any,
  title={“Any bit of help, helps”: Understanding how older caregivers use carework platforms for caregiving support},
  author={Li, Lin and Arnold, Vitica and Piper, Anne Marie},
  booktitle={Proceedings of the 2023 CHI Conference on Human Factors in Computing Systems},
  pages={1--17},
  year={2023}
}

@article{manthorpe2016person,
  title={Person-centered dementia care: current perspectives},
  author={Manthorpe, Jill and Samsi, Kritika},
  journal={Clinical interventions in aging},
  pages={1733--1740},
  year={2016},
  publisher={Taylor \& Francis}
}

@article{postma2015automatic,
  title={Automatic detection of confusion in elderly users of a web-based health instruction video},
  author={Postma-Nilsenov{\'a}, Marie and Postma, Eric and Tates, Kiek},
  journal={Telemedicine and e-Health},
  volume={21},
  number={6},
  pages={514--519},
  year={2015},
  publisher={SAGE Publications Sage CA: Los Angeles, CA}
}

@inproceedings{borges2019classifying,
  title={Classifying confusion: autodetection of communicative misunderstandings using facial action units},
  author={Borges, Niklas and Lindblom, Ludvig and Clarke, Ben and Gander, Anna and Lowe, Robert},
  booktitle={2019 8th International Conference on Affective Computing and Intelligent Interaction Workshops and Demos (ACIIW)},
  pages={401--406},
  year={2019},
  organization={IEEE}
}

@article{sayegh2013cross,
  title={Cross-cultural differences in dementia: the Sociocultural Health Belief Model},
  author={Sayegh, Philip and Knight, Bob G},
  journal={International psychogeriatrics},
  volume={25},
  number={4},
  pages={517--530},
  year={2013},
  publisher={Cambridge University Press}
}

@inproceedings{hsu2023co,
  title={Co-designing social robots with people living with dementia: Fostering identity, connectedness, security, and autonomy},
  author={Hsu, Long-Jing and Bays, Janice K and Tsui, Katherine M and Sabanovic, Selma},
  booktitle={Proceedings of the 2023 ACM Designing Interactive Systems Conference},
  pages={2672--2688},
  year={2023}
}

@article{kamran2022web,
  title={Web simplification prototype for cognitive disabled users},
  author={Kamran, Maira and Malik, Marium and Iqbal, Muhammad Waseem and Anwar, Muhammad and Aqeel, Muhammad and Ahmad, Sana},
  journal={Human Behavior and Emerging Technologies},
  volume={2022},
  number={1},
  pages={5817410},
  year={2022},
  publisher={Wiley Online Library}
}

@article{vendrell2026scaffolding,
  title={Scaffolding critical thinking with generative AI: Design principles for integrating large language models in higher education},
  author={Vendrell, Mireia and Johnston, Samantha-Kaye},
  journal={Computers and Education: Artificial Intelligence},
  pages={100572},
  year={2026},
  publisher={Elsevier}
}

@article{moreno2024designing,
  title={Designing user interfaces for content simplification aimed at people with cognitive impairments},
  author={Moreno, Lourdes and Petrie, Helen and Mart{\'\i}nez, Paloma and Alarcon, Rodrigo},
  journal={Universal access in the information society},
  volume={23},
  number={1},
  pages={99--117},
  year={2024},
  publisher={Springer}
}

@inproceedings{weng2026blessing,
  title={A Blessing and a Challenge: Unpacking Boundary Ambiguities Experienced by Caregivers of Older Adults},
  author={Weng, Tzu-Yu and Bhat, Karthik S},
  booktitle={Proceedings of the 2026 CHI Conference on Human Factors in Computing Systems},
  pages={1--18},
  year={2026}
}

@article{smebye2012persons,
  title={How do persons with dementia participate in decision making related to health and daily care? A multi-case study},
  author={Smebye, Kari Lislerud and Kirkevold, Marit and Engedal, Knut},
  journal={BMC health services research},
  volume={12},
  number={1},
  pages={241},
  year={2012},
  publisher={Springer}
}

@article{daly2018shared,
  title={Shared decision-making for people living with dementia in extended care settings: a systematic review},
  author={Daly, Rachel Louise and Bunn, Frances and Goodman, Claire},
  journal={BMJ open},
  volume={8},
  number={6},
  pages={e018977},
  year={2018},
  publisher={British Medical Journal Publishing Group}
}

@article{gui2023enhancing,
  title={Enhancing Caregiver Empowerment Through the Story Mosaic System: Human-Centered Design Approach for Visualizing Older Adult Life Stories},
  author={Gui, Fang and Yang, Jiaoyun and Wu, Qilin and Liu, Yang and Zhou, Jia and An, Ning},
  journal={JMIR aging},
  volume={6},
  number={1},
  pages={e50037},
  year={2023},
  publisher={JMIR Publications Inc., Toronto, Canada}
}

@inproceedings{mentis2019upside,
  title={Upside and downside risk in online security for older adults with mild cognitive impairment},
  author={Mentis, Helena M and Madjaroff, Galina and Massey, Aaron K},
  booktitle={Proceedings of the 2019 CHI Conference on Human Factors in Computing Systems},
  pages={1--13},
  year={2019}
}

@article{mingxi2025can,
  title={Can AI Become a Friend to Older Adults? Exploring Chatbot Interaction Design Strategies to Alleviate Social Isolation},
  author={Mingxi, Sun and Zhifeng, Zhao},
  journal={International Journal of Human--Computer Interaction},
  pages={1--21},
  year={2025},
  publisher={Taylor \& Francis}
}

@inproceedings{shi2026humans,
  title={When Humans Don't Feel Like an Option: Contextual Factors That Shape When Older Adults Turn to Conversational AI for Emotional Support},
  author={Shi, Mengqi and Song, Tianqi and Zhu, Zicheng and Lee, Yi-Chieh},
  booktitle={Proceedings of the Extended Abstracts of the 2026 CHI Conference on Human Factors in Computing Systems},
  pages={1--8},
  year={2026}
}

@inproceedings{tang2025ai,
  title={AI literacy education for older adults: Motivations, challenges and preferences},
  author={Tang, KangJie, Eugene and Song, Tianqi and Zhu, Zicheng and Li, Jingshu and Lee, Yi-Chieh},
  booktitle={Proceedings of the Extended Abstracts of the CHI Conference on Human Factors in Computing Systems},
  pages={1--15},
  year={2025}
}

@article{vinay2025grace,
  title={GRACE, A Hybrid Rule-and LLM-based Embodied Voice Assistant for Cognitive Stimulation in Older Adults: A Pilot Study Assessing Technical Feasibility, Technology Acceptance, and Working Alliance},
  author={Vinay, Rasita and Uetova, Ekaterina and Tommila, Nora Camilla and Biller-Andorno, Nikola and Kowatsch, Tobias},
  journal={JMIR Aging},
  year={2025},
  publisher={JMIR Publications Inc.}
}

@inproceedings{kim2026clarifying,
  title={Clarifying or Complicating?: Understanding Older Adults' Engagement with Real-World XAI in E-Commerce},
  author={Kim, Seo Hyeong and Kim, Esther Hehsun and Yang, Huiyeon and Lee, Joonhwan and Lim, Hajin},
  booktitle={Proceedings of the 2026 CHI Conference on Human Factors in Computing Systems},
  pages={1--19},
  year={2026}
}

@inproceedings{mccarren2026exploring,
  title={Exploring the Design of a LLM-Based AI Assistant for Mindfulness Practice With Older Adults},
  author={McCarren, Lucy and Eriksson, Ulrika and Ortiz Mengual, Laura and Kuoppam{\"a}ki, Sanna},
  booktitle={Proceedings of the 2026 CHI Conference on Human Factors in Computing Systems},
  pages={1--16},
  year={2026}
}

@inproceedings{vetter2026calls,
  title={Calls of Care: Materializing Posthuman Personhood with Conversational Agents in Dementia Care},
  author={Vetter, Ralf and Hirschmanner, Matthias and Dobrosovestnova, Anna and Frauenberger, Christopher},
  booktitle={Proceedings of the 2026 CHI Conference on Human Factors in Computing Systems},
  pages={1--18},
  year={2026}
}

@inproceedings{ko2025we,
  title={" We need to avail ourselves of [GenAI] to enhance knowledge distribution": Empowering Older Adults through GenAI Literacy},
  author={Ko, Eunhye Grace and Nanayakkara, Shaini and Huff Jr, Earl W},
  booktitle={Proceedings of the Extended Abstracts of the CHI Conference on Human Factors in Computing Systems},
  pages={1--7},
  year={2025}
}

@inproceedings{shandilya2022understanding,
  title={Understanding older adults’ perceptions and challenges in using AI-enabled everyday technologies},
  author={Shandilya, Esha and Fan, Mingming},
  booktitle={Proceedings of the Tenth International Symposium of Chinese CHI},
  pages={105--116},
  year={2022}
}

@inproceedings{madjaroff2017narratives,
  title={Narratives of older adults with mild cognitive impairment and their caregivers},
  author={Madjaroff, Galina and Mentis, Helena},
  booktitle={Proceedings of the 19th international ACM SIGACCESS conference on computers and accessibility},
  pages={140--149},
  year={2017}
}

@inproceedings{huang2026collaborative,
  title={Collaborative AI Scaffolding for Structured Drawing in Dementia Care: A Feasibility Study},
  author={Huang, Hui-Lien and Chen, I-Ping},
  booktitle={Proceedings of the Extended Abstracts of the 2026 CHI Conference on Human Factors in Computing Systems},
  pages={1--5},
  year={2026}
}

@inproceedings{guedes2024scaffolding,
  title={Scaffolding for inclusive co-design: supporting people with cognitive and learning disabilities},
  author={Guedes, Leandro S and Zanardi, Irene and Mastrogiuseppe, Marilina and Span, Stefania and Landoni, Monica},
  booktitle={International Conference on Human-Computer Interaction},
  pages={151--170},
  year={2024},
  organization={Springer}
}

@inproceedings{hu2024designing,
  title={Designing scaffolding strategies for conversational agents in dialog task of neurocognitive disorders screening},
  author={Hu, Jiaxiong and Li, Junze and Zeng, Yuhang and Yang, Dongjie and Liang, Danxuan and Meng, Helen and Ma, Xiaojuan},
  booktitle={Proceedings of the 2024 CHI Conference on Human Factors in Computing Systems},
  pages={1--21},
  year={2024}
}

@inproceedings{bircanin2021including,
  title={Including adults with severe intellectual disabilities in co-design through active support},
  author={Bircanin, Filip and Brereton, Margot and Sitbon, Laurianne and Ploderer, Bernd and Azaabanye Bayor, Andrew and Koplick, Stewart},
  booktitle={Proceedings of the 2021 CHI Conference on Human Factors in Computing Systems},
  pages={1--12},
  year={2021}
}

@inproceedings{ppali2025creating,
  title={Creating with Care: Co-Designing Immersive Experiences through Art-Making with People Living with Dementia},
  author={Ppali, Sophia and Cheung, Ethan and Covaci, Alexandra and She, Wan-Jou and Ang, Chee Siang},
  booktitle={Proceedings of the 2025 CHI Conference on Human Factors in Computing Systems},
  pages={1--18},
  year={2025}
}

@article{breithaupt2025designing,
  title={Designing and Evaluating a Conversational Agent for Early Detection of Alzheimer's Disease and Related Dementias},
  author={Breithaupt, Andrew G and Choi, Nayoung and Finch, James D and Powell, Jeanne M and Nelson, Arin L and Alon, Oz A and Rosen, Howard J and Choi, Jinho D},
  journal={arXiv e-prints},
  pages={arXiv--2509},
  year={2025}
}

@inproceedings{arets2026shared,
  title={Shared Stories, Shared Bonds: People with Dementia Exploring Generative AI Together},
  author={Arets, Teis and Houben, Maarten and van Haeren, Fleur and IJsselsteijn, Wijnand and Perugia, Giulia},
  booktitle={Proceedings of the 2026 CHI Conference on Human Factors in Computing Systems},
  pages={1--19},
  year={2026}
}

@inproceedings{kim2022mymove,
  title={Mymove: Facilitating older adults to collect in-situ activity labels on a smartwatch with speech},
  author={Kim, Young-Ho and Chou, Diana and Lee, Bongshin and Danilovich, Margaret and Lazar, Amanda and Conroy, David E and Kacorri, Hernisa and Choe, Eun Kyoung},
  booktitle={Proceedings of the 2022 CHI Conference on Human Factors in Computing Systems},
  pages={1--21},
  year={2022}
}

@inproceedings{xygkou2024mindtalker,
  title={MindTalker: Navigating the complexities of AI-enhanced social engagement for people with early-stage dementia},
  author={Xygkou, Anna and Ang, Chee Siang and Siriaraya, Panote and Kopecki, Jonasz Piotr and Covaci, Alexandra and Kanjo, Eiman and She, Wan-Jou},
  booktitle={Proceedings of the 2024 CHI conference on human factors in computing systems},
  pages={1--15},
  year={2024}
}

@article{wang2025enabling,
  title={Enabling Older Adults to Provide High-quality Activity Labels: Unpacking Accuracy, Precision, and Granularity in Activity Labeling},
  author={Wang, Yiwen and Khayami, Hossein and Lee, Bongshin and Lazar, Amanda and Kacorri, Hernisa and Choe, Eun Kyoung},
  journal={Proceedings of the ACM on Interactive, Mobile, Wearable and Ubiquitous Technologies},
  volume={9},
  number={4},
  pages={1--24},
  year={2025},
  publisher={ACM New York, NY, USA}
}

@inproceedings{dixon2021taking,
  title={“Taking care of myself as long as I can”: How people with dementia configure self-management systems},
  author={Dixon, Emma and Piper, Anne Marie and Lazar, Amanda},
  booktitle={Proceedings of the 2021 CHI conference on human factors in computing systems},
  pages={1--14},
  year={2021}
}

@inproceedings{berridge2022control,
  title={Control matters in elder care technology: Evidence and direction for designing it in},
  author={Berridge, Clara and Zhou, Yuanjin and Lazar, Amanda and Porwal, Anupreet and Mattek, Nora and Gothard, Sarah and Kaye, Jeffrey},
  booktitle={Proceedings of the 2022 ACM Designing Interactive Systems Conference},
  pages={1831--1848},
  year={2022}
}

@article{treder2024introduction,
  title={Introduction to Large Language Models (LLMs) for dementia care and research},
  author={Treder, Matthias S and Lee, Sojin and Tsvetanov, Kamen A},
  journal={Frontiers in dementia},
  volume={3},
  pages={1385303},
  year={2024},
  publisher={Frontiers Media SA}
}

@inproceedings{dai2021surfacing,
  title={Surfacing the voices of people with dementia: Strategies for effective inclusion of proxy stakeholders in qualitative research},
  author={Dai, Jiamin and Moffatt, Karyn},
  booktitle={Proceedings of the 2021 CHI conference on human factors in computing systems},
  pages={1--13},
  year={2021}
}

@inproceedings{houben2022designing,
  title={Designing for everyday sounds at home with people with dementia and their partners},
  author={Houben, Maarten and Brankaert, Rens and Kenning, Gail and Bongers, Inge and Eggen, Berry},
  booktitle={Proceedings of the 2022 CHI Conference on Human Factors in Computing Systems},
  pages={1--15},
  year={2022}
}

@inproceedings{dixon2020role,
  title={The role of sensory changes in everyday technology use by people with mild to moderate dementia},
  author={Dixon, Emma and Lazar, Amanda},
  booktitle={Proceedings of the 22nd International ACM SIGACCESS Conference on Computers and Accessibility},
  pages={1--12},
  year={2020}
}

@article{vines2015age,
  title={An age-old problem: Examining the discourses of ageing in HCI and strategies for future research},
  author={Vines, John and Pritchard, Gary and Wright, Peter and Olivier, Patrick and Brittain, Katie},
  journal={ACM Transactions on Computer-Human Interaction (TOCHI)},
  volume={22},
  number={1},
  pages={1--27},
  year={2015},
  publisher={ACM New York, NY, USA}
}

@article{pradhan2020use,
  title={Use of intelligent voice assistants by older adults with low technology use},
  author={Pradhan, Alisha and Lazar, Amanda and Findlater, Leah},
  journal={ACM Transactions on Computer-Human Interaction (TOCHI)},
  volume={27},
  number={4},
  pages={1--27},
  year={2020},
  publisher={ACM New York, NY, USA}
}

@inproceedings{lazar2016designing,
  title={Designing for the third hand: Empowering older adults with cognitive impairment through creating and sharing},
  author={Lazar, Amanda and Cornejo, Raymundo and Edasis, Caroline and Piper, Anne Marie},
  booktitle={Proceedings of the 2016 ACM Conference on Designing Interactive Systems},
  pages={1047--1058},
  year={2016}
}

@inproceedings{vidas2024wouldn,
  title={" I wouldn't like technology to give people a get out of jail free card": Caregiver Reflections on a Hypothetical AI-assisted Music Intervention for Dementia Care},
  author={Vidas, Dianna and Kelly, Ryan M and Waycott, Jenny and Thompson, Zara and Tamplin, Jeanette and Kulik, Lars and Vieira Sousa, Tanara and Baker, Felicity A},
  booktitle={Proceedings of the 36th Australasian Conference on Human-Computer Interaction},
  pages={490--499},
  year={2024}
}

@inproceedings{piper2016technological,
  title={Technological caregiving: Supporting online activity for adults with cognitive impairments},
  author={Piper, Anne Marie and Cornejo, Raymundo and Hurwitz, Lisa and Unumb, Caitlin},
  booktitle={Proceedings of the 2016 chi conference on human factors in computing systems},
  pages={5311--5323},
  year={2016}
}

@inproceedings{king2024safespace,
  title={SafeSpace, the smart caretaker: An AI-driven safe and comfortable environment for the well-being of Alzheimer’s and dementia patients},
  author={King III, Kenneth and Azab, Mohamed},
  booktitle={International Conference on Human-Computer Interaction},
  pages={358--369},
  year={2024},
  organization={Springer}
}

@inproceedings{khot2026temporal,
  title={Temporal Snapshots: Probing and Designing for Subjective Time in Dementia},
  author={Khot, Rucha and Brankaert, Rens and IJsselsteijn, Wijnand and Lee, Minha},
  booktitle={Proceedings of the 2026 CHI Conference on Human Factors in Computing Systems},
  pages={1--21},
  year={2026}
}

@inproceedings{baldauf2018exploring,
  title={Exploring requirements and opportunities of conversational user interfaces for the cognitively impaired},
  author={Baldauf, Matthias and B{\"o}sch, Raffael and Frei, Christian and Hautle, Fabian and Jenny, Marc},
  booktitle={Proceedings of the 20th International Conference on human-computer interaction with mobile devices and services adjunct},
  pages={119--126},
  year={2018}
}

@article{ishibashi2025preference,
  title={Preference-Aligned Options from Generative AI Compensates for Age-Related Cognitive Decline in Decision Making},
  author={Ishibashi, Sayaka and Tamura, Kou and Goma, Ayana and Yamamoto, Kenta and Masumoto, Kouhei},
  journal={arXiv preprint arXiv:2511.21164},
  year={2025}
}

@inproceedings{mathur2026wants,
  title={" Who wants to be nagged by AI?": Investigating the Effects of Agreeableness on Older Adults' Perception of LLM-Based Voice Assistants' Explanations},
  author={Mathur, Niharika and Rahman, Hasibur and Desai, Smit},
  booktitle={Proceedings of the Extended Abstracts of the 2026 CHI Conference on Human Factors in Computing Systems},
  pages={1--6},
  year={2026}
}

@inproceedings{mathur2026sometimes,
  title={Sometimes You Need Facts, and Sometimes a Hug: Understanding Older Adults' Preferences for Explanations in LLM-Based Conversational AI Systems},
  author={Mathur, Niharika and Zubatiy, Tamara and Rozga, Agata and Forlizzi, Jodi and D Mynatt, Elizabeth},
  booktitle={Proceedings of the 2026 CHI Conference on Human Factors in Computing Systems},
  pages={1--21},
  year={2026}
}

@inproceedings{pradhan2023towards,
  title={Towards a System Architecture for Connected Physical and Digital Reminders Using Embodied Objects for People with Dementia},
  author={Pradhan, Alisha and Gallier, James and Domjan, Raymond and Dixon, Emma and Bao, Robert and Maddali, Hanuma Teja and Lazar, Amanda},
  booktitle={Proceedings of the 25th International ACM SIGACCESS Conference on Computers and Accessibility},
  pages={1--5},
  year={2023}
}

@inproceedings{gilman2024training,
  title={Training adults with mild to moderate dementia in ChatGPT: exploring best practices},
  author={Gilman, Elizabeth S and Kot, Sushant and Engineer, Margi and Dixon, Emma},
  booktitle={Companion Proceedings of the 29th International Conference on Intelligent User Interfaces},
  pages={101--106},
  year={2024}
}

@inproceedings{dixon2022mobile,
  title={Mobile phone use by people with mild to moderate dementia: uncovering challenges and identifying opportunities: mobile phone use by people with mild to moderate dementia},
  author={Dixon, Emma and Michaels, Rain and Xiao, Xiang and Zhong, Yu and Clary, Patrick and Narayanan, Ajit and Brewer, Robin N and Lazar, Amanda},
  booktitle={Proceedings of the 24th International ACM SIGACCESS Conference on Computers and Accessibility},
  pages={1--16},
  year={2022}
}

@article{saha2025ai,
  title={Ai vs. humans for online support: Comparing the language of responses from llms and online communities of alzheimer’s disease},
  author={Saha, Koustuv and Jain, Yoshee and Liu, Chunyu and Kaliappan, Sidharth and Karkar, Ravi},
  journal={ACM Transactions on Computing for Healthcare},
  year={2025},
  publisher={ACM New York, NY}
}

@article{kot2026exploring,
  title={Exploring the Role of Generative AI in Dementia Resilience Building Activities: Uncovering Opportunities and Challenges},
  author={Kot, Sushant and Engineer, Margi and Gilman, Elizabeth and Flathmann, Christopher and Pradhan, Alisha and Dixon, Emma},
  journal={ACM Transactions on Computer-Human Interaction},
  volume={33},
  number={1},
  pages={1--45},
  year={2026},
  publisher={ACM New York, NY}
}

@article{boehmer2025too,
  title={Too hard to handle: empowering people with amnestic mild cognitive impairment through innovative human--computer interaction and innovative interfaces},
  author={Boehmer, Martin and Massler, Philip and Kuehnel, Stephan and Damarowsky, Johannes and Sackmann, Stefan},
  journal={Behaviour \& Information Technology},
  pages={1--29},
  year={2025},
  publisher={Taylor \& Francis}
}

@inproceedings{chang2008context,
  title={A context aware handheld wayfinding system for individuals with cognitive impairments},
  author={Chang, Yao-Jen and Tsai, Shih-Kai and Wang, Tsen-Yung},
  booktitle={Proceedings of the 10th international ACM SIGACCESS conference on Computers and accessibility},
  pages={27--34},
  year={2008}
}

@inproceedings{keates2009cognitive,
  title={Cognitive impairments, HCI and daily living},
  author={Keates, Simeon and Kozloski, James and Varker, Philip},
  booktitle={International Conference on Universal Access in Human-Computer Interaction},
  pages={366--374},
  year={2009},
  organization={Springer}
}

@article{clarke2017thematic,
  title={Thematic analysis},
  author={Clarke, Victoria and Braun, Virginia},
  journal={The journal of positive psychology},
  volume={12},
  number={3},
  pages={297--298},
  year={2017},
  publisher={Taylor \& Francis}
}

@article{10.1145/3386296.3386303,
author = {Tanis, Emily Shea and Lewis, Clayton},
title = {Artificial intelligence and the dignity of risk},
year = {2020},
issue_date = {October 2019},
publisher = {Association for Computing Machinery},
address = {New York, NY, USA},
number = {125},
issn = {1558-2337},
url = {https://doi.org/10.1145/3386296.3386303},
doi = {10.1145/3386296.3386303},
journal = {SIGACCESS Access. Comput.},
month = mar,
articleno = {7},
numpages = {1}
}

@article{braun2006using,
  title={Using thematic analysis in psychology},
  author={Braun, Virginia and Clarke, Victoria},
  journal={Qualitative Research in Psychology},
  volume={3},
  number={2},
  pages={77--101},
  year={2006},
  publisher={Taylor \& Francis},
  doi={10.1191/1478088706qp063oa}
}

@article{braun2021thematic,
  title={One size fits all? What counts as quality practice in (reflexive) thematic analysis?},
  author={Braun, Virginia and Clarke, Victoria},
  journal={Qualitative Research in Psychology},
  volume={18},
  number={3},
  pages={328--352},
  year={2021},
  publisher={Taylor \& Francis},
  doi={10.1080/14780887.2020.1769238}
}

\clearpage
\onecolumn
\appendix

\section{Caregiver Demographics}
\label{appendix:1}
Table~\ref{tab:caregivers} reports the demographic characteristics of the 14 caregivers who participated as proxy informants, together with the demographic characteristics of the patients they represented. Demographic characteristics of the 31 directly interviewed patients are reported in Table~\ref{tab:patients} in Section~\ref{sec:participants}.

\begin{table*}[!ht]
\centering
\caption{Demographic Information of Caregivers and Their Care Recipients. U/R: Urban/Rural.}
\label{tab:caregivers}
\setlength{\tabcolsep}{4pt}
\begin{tabularx}{\linewidth}{c c c c c c Y c c c Y}
\toprule
\makecell{Caregiver\\ID} & Role & Age & Gender & U/R & Education &
\makecell[l]{Care \\Recipient\\Severity} &
\makecell{Care \\Recipient\\Age} &
\makecell{Care \\Recipient\\Gender} &
\makecell{Care \\Recipient\\U/R} &
\makecell[l]{Care Recipient\\Education} \\
\midrule
G1  & Doctor & 45 & M & U & PhD         & Moderate & 66 & F & U & Primary school \\
G2  & Family & 56 & F & R & Bachelor    & Moderate & 66 & M & R & High school \\
G3  & Doctor & 55 & M & U & PhD         & Severe   & 66 & F & U & Primary school \\
G4  & Family & 56 & M & U & Bachelor    & Severe   & 67 & F & U & Semi-literate \\
G5  & Doctor & 57 & F & R & PhD         & Severe   & 60 & M & R & Junior high school \\
G6  & Doctor & 47 & M & R & PhD         & Severe   & 71 & F & R & Junior high school \\
G7  & Doctor & 47 & M & R & PhD         & Severe   & 73 & F & R & Primary school \\
G8  & Family & 45 & M & U & Bachelor    & Moderate & 68 & M & U & Junior high school \\
G9  & Doctor & 43 & F & U & PhD         & Severe   & 67 & M & U & Junior high school \\
G10 & Family & 39 & F & R & Bachelor    & Severe   & 65 & F & R & Primary school \\
G11 & Family & 55 & M & R & Bachelor    & Severe   & 63 & M & R & Junior high school \\
G12 & Family & 53 & F & U & High school & Severe   & 72 & M & U & Junior high school \\
G13 & Family & 54 & M & R & Bachelor    & Severe   & 70 & M & R & Junior high school \\
G14 & Family & 54 & F & U & Bachelor    & Severe   & 70 & F & U & Semi-literate \\
\bottomrule
\end{tabularx}%
\end{table*}

\FloatBarrier

\section{Interview Guides}
\label{appendix:2}
Table~\ref{tab:patient_interview} lists the semi-structured interview guide used with patients, and Table~\ref{tab:caregiver_interview} lists the corresponding guide used with caregivers. Both guides were used flexibly: questions were adapted, reordered, or skipped depending on each participant's communication ability and the flow of the conversation.

\begin{table*}[t]
\centering
\caption{Semi-Structured Interview Guide for Patients}
\label{tab:patient_interview}
\setlength{\tabcolsep}{5pt}
\begin{tabular}{p{2.5cm} p{4.5cm} p{6.2cm}}
\toprule
\textbf{Theme/Topic} & \textbf{Core Question} & \textbf{Possible Probes} \\
\midrule

AI Exposure
& What kinds of AI-powered systems or devices related to your health or daily life do you currently use or encounter? 
& Mobile apps, wearables, hospital systems, conversational AI; frequency of use; which are used regularly \\

Usage Context 
& In what situations do you usually use these AI systems? 
& When feeling unwell, forgetting things, seeking information, daily routines, or casual interaction \\

First AI Encounter
& What do you remember about your first experience using these AI systems? 
& Who introduced it, why you started using it, whether it was easy or difficult, whether someone helped \\

AI Support Forms
& How do these AI systems usually provide support or assistance? 
& Reminders, suggestions, explanations, asking questions, asking you to confirm, suggesting contacting others \\

AI Interaction Patterns
& How do these forms of AI support typically occur during use? 
& System-initiated vs. user-initiated; consistent vs. context-dependent \\

Human-AI Dynamics 
& Are there situations where other people are involved when you use these AI systems? 
& Family members, doctors; when and how they become involved; system-triggered vs. user-initiated \\

AI Output Style
& How do these AI systems usually present information to you? 
& Direct answers, step-by-step guidance, multiple options, uncertainty expressions, redirection to others \\

Interpreting AI
& When the AI system provides information or suggestions, how do you usually make sense of it? 
& Whether you reflect, follow directly, double-check, or feel uncertain \\

AI Behavior Rules
& Have you noticed any recurring ways or "rules" in how the AI system works? 
& Repeated questioning, refusal to answer, requiring confirmation, suggesting stopping, risk reminders \\

AI Constraint Impact
& How do these AI behaviors affect your experience of using the system? 
& Easier or harder to use, more reassuring or frustrating, increased or reduced willingness to use \\

Autonomy \& AI
& In your experience, which decisions feel like your own, and which do not? 
& Situations where you decide independently vs. involving others or following AI suggestions \\

AI \& Relationships
& Has using these AI systems changed how you interact with family members or doctors? 
& More frequent consultation, reduced communication, increased reliance \\

Evolving AI Use 
& Has your way of using these AI systems changed over time? 
& Increased familiarity, reliance, reduced effort, selective use \\

Overall AI Reflection 
& Overall, how do you feel about these AI systems? 
& Most helpful aspects, most challenging aspects, suggestions for improvement \\

\bottomrule
\end{tabular}
\end{table*}

\begin{table*}[t]
\centering
\caption{Semi-Structured Interview Guide for Caregivers}
\label{tab:caregiver_interview}
\setlength{\tabcolsep}{5pt}
\begin{tabular}{p{2.5cm} p{4.5cm} p{6.2cm}}
\toprule
\textbf{Theme/Topic} & \textbf{Core Question} & \textbf{Possible Probes} \\
\midrule

Observed AI Use 
& What kinds of AI-powered systems or tools does the patient use in daily life? 
& Health apps, devices, conversational AI systems; frequency and patterns of use \\

AI Usage Situations 
& In what situations do they tend to use these AI systems? 
& Seeking health information, reminders, decision-making, companionship \\

AI Interaction Patterns 
& How do these AI systems typically interact with the patient? 
& Providing answers, asking questions, requiring confirmation, suggesting contacting others \\

AI Design vs. Practice
& Do these forms of AI support come from system design, or do they emerge in practice? 
& Default AI behaviors, institutional settings, family involvement \\

Caregiver \& AI Use
& In what situations do you become involved when the patient uses these AI systems? 
& Helping interpretation, making decisions, operating the system, verifying AI outputs \\

AI Involvement Triggers
& How is your involvement typically introduced? 
& Triggered by AI prompts, patient requests, or proactive intervention \\

Patient AI Comprehension
& How do you think the patient understands the AI system's information or suggestions? 
& Common misunderstandings, limitations, reliance patterns \\

AI Behavior Impact
& What changes in the patient's behavior have you observed when using these AI systems? 
& Increased reliance, reduced initiative, more consultation, changes in interaction patterns \\

AI \& Care Relation
& Have these AI systems changed your interaction with the patient? 
& Redistribution of responsibility, increased or decreased workload \\

Evaluating AI Practices 
& How do you view the recurring practices or constraints in these AI systems? 
& Whether they are appropriate, excessive, or insufficient \\

AI \& Severity Levels
& Do you think these AI practices are equally suitable for different levels of cognitive impairment? 
& Differences across mild, moderate, severe conditions \\

AI Pros and Cons
& From your perspective, what are the main benefits and drawbacks of these AI systems? 
& Safety, controllability, usability, burden, dependency \\

AI Role Over Time
& How has the role of these AI systems changed as the patient's condition evolves? 
& Increasing or decreasing usefulness, need for more intervention \\

Overall AI Evaluation
& Overall, how do you evaluate these AI systems? 
& Areas of AI support, limitations, suggestions for improvement \\

\bottomrule
\end{tabular}
\end{table*}

\end{document}